\documentclass[aps,prd,reprint,amsmath,amssymb,showkeys,nofootinbib]{revtex4-2}
\usepackage{graphicx}	
\usepackage{color}
\usepackage{mathrsfs}
\usepackage[version=4]{mhchem}
\usepackage{booktabs}
\usepackage{hyperref}
\hypersetup{pdfborder = {0 0 0},colorlinks=true,linkcolor=[RGB]{40,43,131},citecolor=[RGB]{40,43,131},urlcolor=[RGB]{40,43,131}}

\newcommand{\RNum}[1]{\uppercase\expandafter{\romannumeral #1\relax}} 

\newcommand{\msun}{M_\odot}

\makeatletter
\let\jnl@style=\rmfamily 
\def\ref@jnl#1{{\jnl@style#1}}%
\newcommand\aj{\ref@jnl{AJ}}
\newcommand\psj{\ref@jnl{PSJ}}
\newcommand\araa{\ref@jnl{ARA\&A}}
\renewcommand\apj{\ref@jnl{ApJ}}
\newcommand\apjl{\ref@jnl{ApJL}}     
\newcommand\apjs{\ref@jnl{ApJS}}
\renewcommand\ao{\ref@jnl{ApOpt}}
\newcommand\apss{\ref@jnl{Ap\&SS}}
\newcommand\aap{\ref@jnl{A\&A}}
\newcommand\aapr{\ref@jnl{A\&A~Rv}}
\newcommand\aaps{\ref@jnl{A\&AS}}
\newcommand\azh{\ref@jnl{AZh}}
\newcommand\baas{\ref@jnl{BAAS}}
\newcommand\icarus{\ref@jnl{Icarus}}
\newcommand\jaavso{\ref@jnl{JAAVSO}}  
\newcommand\jrasc{\ref@jnl{JRASC}}
\newcommand\memras{\ref@jnl{MmRAS}}
\newcommand\mnras{\ref@jnl{MNRAS}}
\renewcommand\pra{\ref@jnl{PhRvA}}
\renewcommand\prb{\ref@jnl{PhRvB}}
\renewcommand\prc{\ref@jnl{PhRvC}}
\renewcommand\prd{\ref@jnl{PhRvD}}
\renewcommand\pre{\ref@jnl{PhRvE}}
\renewcommand\prl{\ref@jnl{PhRvL}}
\newcommand\pasp{\ref@jnl{PASP}}
\newcommand\pasj{\ref@jnl{PASJ}}
\newcommand\qjras{\ref@jnl{QJRAS}}
\newcommand\skytel{\ref@jnl{S\&T}}
\newcommand\solphys{\ref@jnl{SoPh}}
\newcommand\sovast{\ref@jnl{Soviet~Ast.}}
\newcommand\ssr{\ref@jnl{SSRv}}
\newcommand\zap{\ref@jnl{ZA}}
\renewcommand\nat{\ref@jnl{Nature}}
\newcommand\iaucirc{\ref@jnl{IAUC}}
\newcommand\aplett{\ref@jnl{Astrophys.~Lett.}}
\newcommand\apspr{\ref@jnl{Astrophys.~Space~Phys.~Res.}}
\newcommand\bain{\ref@jnl{BAN}}
\newcommand\fcp{\ref@jnl{FCPh}}
\newcommand\gca{\ref@jnl{GeoCoA}}
\newcommand\grl{\ref@jnl{Geophys.~Res.~Lett.}}
\renewcommand\jcp{\ref@jnl{JChPh}}
\newcommand\jgr{\ref@jnl{J.~Geophys.~Res.}}
\newcommand\jqsrt{\ref@jnl{JQSRT}}
\newcommand\memsai{\ref@jnl{MmSAI}}
\newcommand\nphysa{\ref@jnl{NuPhA}}
\newcommand\physrep{\ref@jnl{PhR}}
\newcommand\physscr{\ref@jnl{PhyS}}
\newcommand\planss{\ref@jnl{Planet.~Space~Sci.}}
\newcommand\procspie{\ref@jnl{Proc.~SPIE}}

\newcommand\actaa{\ref@jnl{AcA}}
\newcommand\caa{\ref@jnl{ChA\&A}}
\newcommand\cjaa{\ref@jnl{ChJA\&A}}
\newcommand\jcap{\ref@jnl{JCAP}}
\newcommand\na{\ref@jnl{NewA}}
\newcommand\nar{\ref@jnl{NewAR}}
\newcommand\pasa{\ref@jnl{PASA}}
\newcommand\rmxaa{\ref@jnl{RMxAA}}

\newcommand\maps{\ref@jnl{M\&PS}}
\newcommand\aas{\ref@jnl{AAS Meeting Abstracts}}
\newcommand\dps{\ref@jnl{AAS/DPS Meeting Abstracts}}
\makeatother

\begin{document}
\defcitealias{Li2023}{L23}
\defcitealias{Li2024}{L24}

\title{Gravitational Wave Forecasts Constrained by JWST AGN Observations for\\Early Massive Black Hole Mergers}

\author{Hanpu Liu}
\email{hanpu.liu@princeton.edu}
\affiliation{Kavli Institute for Astronomy and Astrophysics, Peking University, Beijing 100871, China}
\affiliation{Department of Astrophysical Sciences, Princeton University, Princeton, NJ 08544, USA}

\author{Kohei Inayoshi}
\email{inayoshi@pku.edu.cn}
\affiliation{Kavli Institute for Astronomy and Astrophysics, Peking University, Beijing 100871, China}

\begin{abstract}
Massive black holes (BHs) grow by gas accretion and mergers, observable through electromagnetic and gravitational wave (GW) emission. The James Webb Space Telescope (JWST) has detected faint active galactic nuclei (AGNs) powered by accreting BHs with masses of $M_{\bullet}\sim 10^{6-8}~M_\odot$, revealing a previously unknown, abundant population of BHs. This mass range overlaps with the detection scopes of space-based GW interferometers and approaches the upper bounds of the predicted mass of seed BHs. We model BH mass assembly in light of the new JWST findings to investigate their formation channels and predict merger events. Two types of seed BHs are considered: heavy seeds ($M_\bullet \sim 10^{2-5}~M_\odot$) formed in rare and overdense cosmic regions, and light seeds ($M_\bullet \sim 10^{1-3}~M_\odot$) formed as stellar remnants in less massive dark-matter halos. 
The BHs grow through episodic accretion and merger events, which we model by fitting the AGN luminosity function to observational data including JWST-identified AGNs at $z\sim 5$. We find that heavy seeds alone struggle to explain quasars and faint JWST-selected AGNs simultaneously, requiring the more abundant light seeds. The observed merger rate of BHs from heavy seeds alone is limited to $\lesssim 10^{-1}~{\rm yr}^{-1}$ for major mergers at $z\geq5$. However, the presence of light seeds increases the major merger rate by several orders of magnitude, which peaks at a total BH mass of $\simeq 2\times 10^3~M_\odot$ over $5<z<10$ at a rate of $\sim 30~{\rm yr}^{-1}$. These events are detectable by future GW observatories such as the Laser Interferometer Space Antenna (LISA) and the pathfinder to DECi-hertz Interferometer Gravitational-wave Observatory (B-DECIGO). Precise sky localization and distance measurement of those GW events, with solid angle and luminosity distance uncertainties $\Delta\Omega\Delta\log D_L\lesssim 10^{-4}~\rm deg^2$, will enable electromagnetic identification of mergers at $z\geq5$ and subsequent multi-messenger follow-up observations.
\end{abstract}


\maketitle

\vspace{10mm}

\section{Introduction}
Gravitational wave (GW) observations have ushered in a new era for characterizing the cosmic black hole (BH) population.  Ground-based observatories, sensitive to GW frequencies of $\sim10^2~$Hz, now routinely detect mergers involving stellar-mass BHs (mass $M_\bullet\sim10$--$100~M_\odot$, \cite{GWTC32023}, and references therein). Pulsar Timing Array (PTA) experiments present evidence for a stochastic GW background in the nano-Hz (nHz) band, which is ascribed to extremely massive BHs ($M_\bullet>10^9~M_\odot$, \cite{Agazie2023,Reardon2023,EPTA2023,Xu2023}).

Space-based GW interferometers planned in the next decade, e.g., the Laser Interferometer Space Antenna (LISA, \cite{AmaroSeoane2017,AmaroSeoane2023}), the pathfinder to DECi-hertz Interferometer Gravitational-wave Observatory (B-DECIGO, \cite{Kawamura2019}), TianQin \cite{Luo2016}, and Taiji \cite{Luo2020}, will probe GWs in the milli-Hz to deci-Hz (mHz--dHz) bands with advanced sensibilities, characterizing BHs in the intermediate mass range ($M_\bullet \sim 10^{2-7}~M_\odot$) at redshifts up to $z\sim20$. This will chart the mass assembly history of supermassive BHs hosted by massive galaxies in the local universe. Born as seeds in the mass range $10^2~M_\odot\lesssim M_\bullet\lesssim10^5~M_\odot$ at high redshift (see Ref.~\cite{Inayoshi2020} for a review), these BHs are believed to gain mass over cosmic time by accreting surrounding gas and by merging with other BHs \cite{Lynden-Bell1969,Yu2002}.

This evolutionary history is currently probed by electromagnetic (EM) observations when massive BHs (MBHs) rapidly accrete gas and appear as active galactic nuclei (AGN). Thus far, wide-field quasar surveys have identified BHs with masses $M_\bullet\sim10^{8-10}~\msun$ at redshifts $z\sim4\text{--}8$ (e.g., \cite{Akiyama2018,Niida2020,Jiang2016,Matsuoka2018,Schindler2023}). Recently, the James Webb Space Telescope (JWST) has been unveiling faint AGNs at $z\sim4$--7 with abundances one to two orders of magnitude higher than extrapolations of ground-based quasar luminosity functions (LFs; \cite{Onoue2023,Harikane2023,Kocevski2023,Maiolino2023,Matthee2024}). These faint AGNs likely indicate MBHs with less extreme masses, and, together with the quasars, provide increasingly representative statistics to constrain MBH growth models.

Current EM observations of MBHs have yet to extend significantly beyond $z>10$, the epoch when MBHs were seeded and initiated their early growth. Extensive theoretical research has discussed two MBH formation channels. ``Light-seed'' BHs are expected to form abundantly as remnants of the first generation of stars (Population~\RNum{3}, hereafter Pop~\RNum{3}). These Pop~\RNum{3} stars are conceived in small dark-matter halos $(M_{\rm halo}\sim10^{5\text{--}6}~M_\odot)$ at $z\sim20\text{--}30$ \cite{Abel2002,Yoshida2008}, with typical stellar masses up to $10^{2\text{--}3}~M_\odot$ in simulations \cite{Bromm2002,Hirano2014,Stacy2016,Jaura2022}. MBHs seeded from the Pop~\RNum{3} stars likely encounter unfavorable conditions for rapid mass assembly \cite{Kitayama2004,Whalen2004,Johnson2007,Alvarez2009,Tanaka2012} and have difficulty explaining the brightest quasars at $z\sim6$ \cite{Inayoshi2016,Pacucci2017}, although their contribution to the less extreme population remains poorly understood. 

On the other hand, in the ``heavy-seed'' scenario, a massive gas cloud directly collapses into a seed BH of mass $10^{4\text{--}6}~M_\odot$ \cite{Bromm2003,Lodato2006,Shang2010}. Peculiar environmental effects are required to keep the cloud from vigorous fragmentation \cite{Omukai2001,Tanaka2014,Inayoshi2014,Becerra2015,Wise2019}, and these conditions are considered to be too stringent to be commonly realized in typical regions of the high-redshift universe. Recent studies found that progenitors of high-redshift quasar host halos, i.e., rare, overdense cosmic regions, fulfill the physical conditions to form heavy-seed BHs at $z>10$ due to dark-matter halo clustering \cite{Li2021,Lupi2021}. Subsequent mass growth of these BHs explains the $z\gtrsim4$ quasar LFs \cite{Li2023,Li2024,Trinca2022,Jeon2024}.

Connecting seeding models with high-redshift AGN data represents a crucial discovery space for future GW observations and is the focus of this paper. In this work, we extend the model developed by Ref.~\cite{Li2023}, which outlined the episodic accretion history of MBHs from initial seeding through emergence as high-redshift quasars. We incorporate mergers into this framework to predict GW events. Moreover, new JWST data extends constraints on the faint end of the LF, enabling us to examine the necessity of light-seed BHs and thus explore the BH mergers within this previously hidden population. 
We predict that mergers involving heavy seeds alone will be very infrequent (with a fiducial major merger rate 
$\lesssim10^{-1}~{\rm yr}^{-1}$ in the observer's frame), while light-seed BH mergers are expected to occur at significantly higher 
rates of $\sim 30~{\rm yr}^{-1}$. 
LISA is anticipated to detect more than half of these events, whereas B-DECIGO will capture almost all due to its sensitivity to higher frequencies. These observations will offer new insights into the early assembly of MBHs.

This paper is organized as follows. In Section~\ref{sec:method}, we describe our methods of MBH seeding, accretion, mergers, data-fitting, and GW event rate prediction. We find in Section~\ref{sec:heavy} that the heavy-seed population alone has difficulty explaining the JWST AGNs and implies infrequent GW events. In Section~\ref{sec:hl}, we demonstrate that the combined heavy- and light-seed populations account for the unobscured AGN LF and produce plentiful GW events. We discuss in Section~\ref{sec:discussion} model extensions related to the time needed before the BHs may merge, the afterglows of GW events, and the MBH mass density evolution. We also compare our model to previous works. We summarize our findings in Section~\ref{sec:summary}. Our predicted unobscured AGN LF, BHMF, and the merger rate distribution are publicly available on GitHub\footnote{\url{https://github.com/hanpu-liu/BHMF\_AGNLF\_data}}. Throughout this work, we adopt the \citet{Planck2016} cosmological parameters, i.e., $\Omega_{\rm m}$ = 0.307, $\Omega_\Lambda$ = 0.693, $\Omega_{\rm b}$ = 0.0486, and $H_0 = 67.7~\rm km\,s^{-1}\,Mpc^{-1}$. All densities in this work are based on the comoving volume. All magnitudes in this work are in the AB system.

\section{Method}
\label{sec:method}
\begin{figure*}
    \centering
    \includegraphics[width=0.98\textwidth]{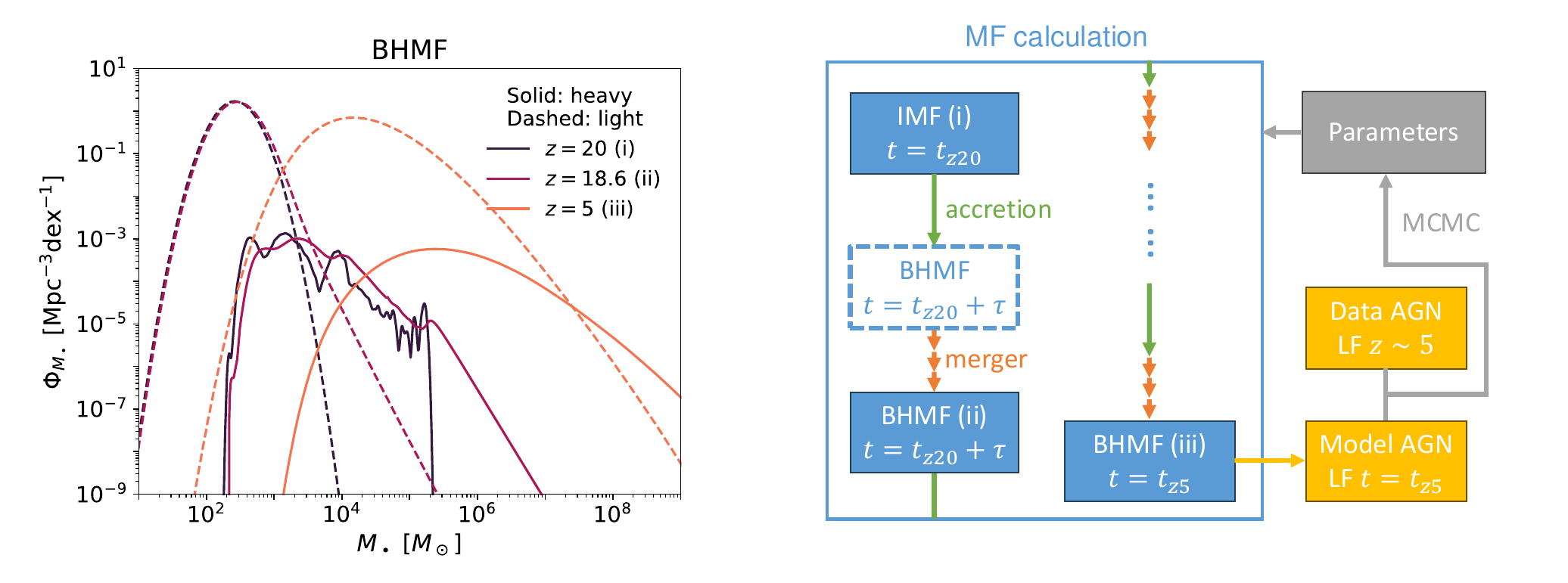}
    \caption{Method overview. \textit{Left}: BHMF at the initial seeding, after the first accretion time step, and at the terminal redshift. \textit{Right}: method flowchart. The IMF (i), BHMF (ii), and BHMF (iii) boxes correspond to the curves on the left panel. The box with dashed borders indicates an intermediate state when accretion during $\tau$ is complete but mergers are not calculated yet.}
    \label{fig:flowchart}
\end{figure*}

Figure~\ref{fig:flowchart} provides an overview of our method. We calculate the BH number distribution per unit comoving volume per logarithmic mass, known as the black hole mass function (BHMF). At birth, the heavy- and light-seed BHs are introduced (Section~\ref{subsec:seeding}), which then evolve via accretion (Section~\ref{subsec:accretion}) and mergers (Section~\ref{subsec:merger_coagulation}) in multiple time steps. The BHMF at $z\sim5$ is converted to the unobscured AGN LF and fit to observational data (Section~\ref{subsec:MCMC}). The calculation is repeated in the Markov chain Monte Carlo (MCMC) method to sample the parameter space. The best-fit parameter set and the corresponding evolutionary history imply the properties of merger events to be detected by GW observatories (Section~\ref{subsec:GW_method}). The cosmic time $t_{z20}=178.5~\rm Myr$ and $t_{z5}=1173~\rm Myr$ corresponds to $z=20$ and $z=5$.

\subsection{The birth of MBHs: heavy and light seeds}
\label{subsec:seeding}
In the framework of structure formation, primordial density fluctuations give rise to dark-matter halos, whose potential wells accumulate baryons that later build galaxies and MBHs. In highly biased, overdense regions at $z\gtrsim10$--20, heavy-seed BHs form in the main progenitors of quasar host galaxies via the collapse of massive baryonic clouds. The cloud is kept warm by \ce{H2}-dissociating radiation, successive dark-matter halo mergers, and baryonic streaming motion until the gas mass accumulated in the halo substantially exceeds the Jeans mass of the warm gas. Since the gas remains nearly isothermal via atomic cooling, the gas cloud collapses without efficient fragmentation and eventually leaves a heavy-seed BH (see \cite{Inayoshi2020} for a review). Based on this, Ref.~\cite{Li2023} modeled the initial mass function (IMF) of heavy-seed BHs with a number density $\sim 10^{-3}~\rm Mpc^{-3}$ and a mass range $2 \times 10^2~\msun \lesssim M_\bullet \lesssim 2\times10^5~\msun$, as shown in Figure~\ref{fig:flowchart}. Note that, from now on, we distinguish ``heavy-seed'' versus ``light-seed'' BHs based on their cosmological origin rather than conventionally on formation mechanisms. The diversity of the progenitor halo properties in overdense environments self-consistently gives rise to BHs from both Pop~\RNum{3} stars and direct collapse, hence the continuous mass span of the heavy-seed IMF in Figure~\ref{fig:flowchart}. 
This is distinct from scenarios that assume a bimodal distribution (e.g., $M_\bullet\sim10^{2\text{--}3}~\msun$ Pop~\RNum{3} remnants vs. $10^{4\text{--}6}~\msun$ direct-collapse BHs) or focus exclusively on one seed population. The vast majority of these BHs have already formed by $z_0=20$, which when evolved to $z\sim6$ explain the quasar LF from ground-based surveys. We use the same seeding process in this work except that we treat $n_{0h}$, the number density of heavy-seed BHs at $z_0$, as a variable parameter. 

In contrast, in commoner minihalos in the early universe, Pop~\RNum{3} stars become the only feasible MBH formation channel. Efficient cooling of molecular hydrogen leads to cloud fragmentation and the formation of the first generation of protostars \cite{Yoshida2008}, which will likely evolve into BHs of several tens or hundreds of solar masses \cite{Yoon2012} in a million-year timescale. These light-seed BHs, formed prior to heavy-seed BHs, may also contribute to the AGN population. However, their accretion will be severely limited by the gas evacuating from the shallow gravitational potential well, until the host halo becomes sufficiently massive to gravitationally bound the gas heated by BH feedback
\cite{Alvarez2009,Jeon2012,Tanaka2012}. Mergers present an alternative mass assembly channel, but GW recoils eject the merging BHs at velocities ($\gtrsim 100~{\rm km~s}^{-1}$, depending on the mass ratio and spin configuration) typically exceeding the escape velocity of minihalos ($\lesssim 10~{\rm km~s}^{-1}$). This likely isolates the descendent BH from the gas or other BHs in halos and suppresses its subsequent growth \cite{Haiman2004}. Therefore, even though Pop~\RNum{3} stars may be ubiquitous in minihaloes, rapid growth of their descendent BHs is feasible only in massive halos with a virial temperature of $T_{\rm vir} \gtrsim 10^4~\rm K$, which provide a dense gas reservoir \cite{Inayoshi2016,Ryu2016} and sufficient gravitational binding to prevent the gas or the BH itself from escaping. 

In this work, we adopt the number density of such atomically-cooling halos with $T_{\rm vir}\simeq 10^4~{\rm K}$ as an upper bound of the number density $n_{0l}$ of light-seed BHs. In the redshift range of initial BH growth, $z\gtrsim 10$, the number density is calculated as $\sim 1-10~\rm Mpc^{-3}$ \cite{Sheth2001}, and thus we consider two cases with $n_{0l}=1$ and $10~\rm Mpc^{-3}$. These values are substantially lower than those estimated in previous studies that focused on the bulk population of Pop~\RNum{3} stars in more abundant minihalos and aimed to explain stellar-mass binary BH formation or chemical enrichment (e.g., $10^2~\rm Mpc^{-3}$ in Ref.~\cite{Tanikawa2021}; see also \cite{Magg2016,Skinner2020}). We insert light-seed BHs in a single event at $z_0=20$ and assume that massive Pop~\RNum{3} stars hardly form later due to the rise of H$_2$-dissociating far-ultraviolet radiation emitted from nearby galaxies \cite{Haiman1997,Machacek2001,Johnson2013}. We adopt a log-normal BHMF centered on $263~M_\odot$ with a log-space standard deviation of 0.235, as shown in Figure~\ref{fig:flowchart}. These parameters are determined by fitting the Pop~\RNum{3} IMF derived from Ref.~\cite{Hirano2015}\footnote{The cited work showed a maximum stellar mass of $\sim10^3~\msun$ due to their limited sample size, while our log-normal distribution allows some light-seed BHs to have masses $>10^3~\msun$. However, the massive end of the light-seed IMF hardly influences our results. We tested by calculating the $z\sim5$ AGN LF with the same best-fit model parameters in our different cases, introduced later, while imposing a maximum of $10^3~\msun$ on the light-seed BH IMF. The goodness of fit (measured in log posterior probability) would only deviate by $<1.0$. }. We neglect mass loss during stellar evolution and pair-instability supernovae
given the uncertainties in the number density and IMF of Pop~\RNum{3} stars (see Ref.~\cite{Klessen2023} for a review). 

\subsection{MBH episodic accretion from initial seeding to $z\sim 5$}
\label{subsec:accretion}
We briefly describe the MBH evolution model due to episodic accretion, similar to that in Ref.~\cite{Li2023}. Starting from initial seeding at $z\sim20$, the BHMF is updated at each accretion episode with a timescale of $10^{6\text{--}8}$ yr. The accretion keeps the total comoving BH number density unchanged but modifies the abundance distribution. 

A minimum number of parameters control accretion. The BH accretion rate is given by
\begin{equation}
    \dot{M}_\bullet = (1-\epsilon)\lambda f(M_\bullet)\dot{M}_{\rm Edd}\,, \label{eq:accretion_rate}
\end{equation}
where $\lambda$ is the ratio of the bolometric luminosity of the accreting SMBH to its Eddington luminosity $L_{\rm Edd}$, and $\dot{M}_{\rm Edd}\equiv L_{\rm Edd}/\epsilon c^2$ is the Eddington accretion rate assuming a radiative efficiency of $\epsilon=0.091$ (\cite{Shakura1973}; the efficiency is consistent with that from the Soltan argument, e.g., \cite{Soltan1982,Yu2002}). The function
\begin{equation}
    f(M_\bullet) = \frac{2}{1 + (M_\bullet / M_{\bullet,\rm c})^\delta}\,, \label{eq:delta}
\end{equation}
where we adopt $M_{\bullet,\rm c}=10^8~M_\odot$ \cite{Ueda2014} and $\delta>0$, characterizes the positive mass-dependent radiative efficiency as found among AGN at $z\lesssim3$ \cite{Cao2008,Li2012}. The growth of the most massive BHs is thereby suppressed, although this effect becomes negligible in the limit of $\delta\ll1$, where $f(M_\bullet)\simeq1$ and Equation~(\ref{eq:accretion_rate}) reduces to exponential growth with an $e$-folding timescale of $M_\bullet/(1-\epsilon)\lambda\dot{M}_{\rm Edd}= (45/\lambda)$ Myr. 

Observation has constrained quasar activity to a cosmologically short lifetime ($\sim10^{6\text{--}8}$~yr, \cite{Martini2004}), consistent with the theoretical picture that accretion bursts with strong gas inflow are limited by BH feedback \cite{DiMatteo2005,Hopkins2005}. This suggests an episodic MBH growth history involving diverse Eddington ratio values. Here, we model the Eddington ratio distribution function (ERDF) as a Schechter function using two free parameters, $\lambda_0$ and $\alpha$:
\begin{equation}
    g(\lambda) \equiv \frac{dP}{d\log\lambda} = \begin{cases} 
    \displaystyle\frac{\ln10}{\Gamma(\alpha,\frac{\lambda_{\rm min}}{\lambda_0})}\left(\frac{\lambda}{\lambda_0}\right)^\alpha e^{-\frac{\lambda}{\lambda_0}}\,, & \lambda \geq \lambda_{\rm min}\,, \\
    0\,, & \text{otherwise\,,}
    \end{cases}
    \label{eq:erdf}
\end{equation}
where the prefactor with the incomplete Gamma function normalizes the probability distribution. The profile fits well the ERDF of low-$z$ AGNs over $0.01\lesssim\lambda\lesssim1$ \cite{Schulze2015}. The exponential cutoff allows for a low but nonzero probability of super-Eddington accretion. Following Ref.~\cite{Li2023}, we adopt the minimum Eddington ratio $\lambda_{\rm min}=0.01$ suggested by X-ray AGN observations \cite{Aird2018} and BH feedback models \cite{Novak2011}. We caution that this minimum is uncertain. A lower value of $\lambda_{\rm min}=10^{-3}$ would significantly increase the probability of a BH being inactive for fixed $\lambda_0$ and $\alpha$. In our fiducial case in Section~\ref{sec:hl}, model--data agreement would then require $\alpha$ to increase from $-0.027$ to $0.118$ to attenuate the low end of the ERDF, while the other best-fit parameter values are less affected.

To model the episodic accretion activity, we introduce a time duration $\tau$, during which the mass growth of an MBH is governed by Equation~(\ref{eq:accretion_rate}) with a fixed $\lambda$ randomly assigned from Equation~(\ref{eq:erdf}). The MBH in its entire evolutionary history undergoes many such episodes with different values of $\lambda$, randomly alternating its accretion rate and luminosity. Thus, the BHMF $\Phi_{M_\bullet}$ at a given time $t$ is updated to $t+\tau$ with
\begin{equation}
    \Phi_{M_\bullet}(M_\bullet,t+\tau) = \int g(\lambda_*)\frac{\partial\log\lambda_*}{\partial\log M_{\bullet}}\Phi_{M_\bullet}(M_{\bullet,0},t)d\log M_{\bullet,0}\,,
    \label{eq:BHMF_evolution_accretion}
\end{equation}
where $\lambda_*(M_\bullet,M_{\bullet,0},\tau)$ is the Eddington ratio required for a BH with $M_{\bullet,0}$ to grow up to $M_\bullet$ in $\tau$, implied by Equation~(\ref{eq:accretion_rate}). Integrating over $\log M_\bullet$ on both sides of Equation~(\ref{eq:BHMF_evolution_accretion}), one finds that the total number densities are the same before and after the accretion. 

We newly introduce the light-seed BH population compared to Ref.~\cite{Li2023}. From now on, we distinguish the heavy- and light-seed quantities with the subscripts $``h"$ and $``l"$ (e.g., the total BHMF is decomposed as $\Phi_{M_\bullet} = \Phi_{M_\bullet,h}+\Phi_{M_\bullet,l}$). We assume the same episodic accretion behavior for heavy- and light-seed BHs except that the ERDF of the latter has a smaller characteristic Eddington ratio, i.e., $\lambda_{0h}/\lambda_{0l}>1$, in line with the theoretical expectation of relatively slow accretion (see Section~\ref{subsec:seeding}). This formulation implies that the light-seed population only accounts for faint AGNs at high redshift, leaving the brighter quasars to the rare, heavy-seed BHs. 

\subsection{MBH mergers with the coagulation model}
\label{subsec:merger_coagulation}
In contrast to accretion, mergers boost the average BH mass at the expense of the total number density. After accretion in each time step $\tau$, reaching the intermediate state in the flowchart in Figure~\ref{fig:flowchart}, we consider mergers by updating the BHMF in the same time step using the coagulation formalism. 

The coagulation model describes the evolution of a population of objects undergoing mergers \cite{Smoluchowski1916}. The theory applies to any physical system involving two-body interactions and has been extensively used in nuclear reactions, asteroids, aerosols, etc. The following assumptions are made. Firstly, the merging BHs are spatially homogeneous, i.e., the merging process is independent of location. Secondly, each merger is a two-body process, whose merger rate is proportional to the product of the abundance of each component. Mergers involving three or more BHs are naively rarer as the merger rate will scale to higher orders of the abundance, but may become a significant orbital decay mechanism for MBH binaries that otherwise do not merge efficiently within a Hubble time (e.g., \cite{HoffmanLoeb2007,Amaro-Seoane2010,Bonetti2019}). Thirdly, individual merger events obey mass conservation, i.e., the merger remnant mass equals the sum of the progenitor masses. We thus ignore mass loss due to GW radiation during mergers. Post-Newtonian calculations indicate that the mass loss relative to the total progenitor mass is $\sim0.05$ for major mergers (mergers involving progenitors of similar masses), depending on the mass ratio and spin (see Equations 4 and 5 in Ref.~\cite{Lousto2010}). Therefore, the coagulation formalism will not significantly overestimate the total mass density unless each MBH on average experiences on the order of 10 major mergers in its evolutionary history, an order of magnitude larger than empirical predictions of galaxy major mergers \cite{OLeary2021}.

Given the above assumptions, the BHMF evolution due to mergers is governed by
\begin{subequations}
    \begin{align}
        \frac{\partial\Phi_{M_\bullet,h}}{\partial t} &= -\phi\Phi_{M_\bullet,h} + (\psi_{hh}+\psi_{hl}+\psi_{lh}) \,, \label{eq:coagulation_h} \\
        \frac{\partial\Phi_{M_\bullet,l}}{\partial t} &= -\phi\Phi_{M_\bullet,l} + \psi_{ll} \,, \label{eq:coagulation_l}
    \end{align}
    \label{eq:coagulation}
\end{subequations}
where
\begin{align}
    \phi(M,t) &=  \int_0^\infty \Phi_{M_\bullet}(m,t) K(M,m,t)~d\log m \,,\nonumber\\
    \psi_{ij}(M,t) & =  M\int_0^{M/2}\Phi_{M_\bullet,i}(m,t)\frac{\Phi_{M_\bullet,j}(M-m,t)}{M-m}\nonumber\\ 
     & \times  K(m,M-m,t)~d\log m \,,\quad i,j=h,l. \label{eq:psi_ij_def}
\end{align}
For a particular infinitesimal (logarithmic) mass bin centered on $M$, the first term on the right-hand side of Equation~(\ref{eq:coagulation}) describes the abundance loss as BHs in the bin participate in mergers, whereas the second term denotes the abundance gain as smaller BHs merge to produce remnants of mass $M$. As we consider two MBH populations, three types of mergers may arise, i.e., heavy-heavy ($\psi_{hh}$), heavy-light ($\psi_{hl}, \psi_{lh}$), and light-light ($\psi_{ll}$). We assume the remnant BH belongs to the heavy-seed category if one or two merger progenitors originate from heavy seeds while the merger between light-seed BHs produces a light-seed one. 
This formulation conserves the BH total mass density defined by
\begin{equation}
    \rho(t)\equiv\int_0^\infty M\Phi_{M_\bullet}(M,t) d\log M\,, \label{eq:mass_density_conservation}
\end{equation}
and thus $\rho$ changes with $t$ only due to accretion.

The coagulation kernel $K(M,m,t)$ characterizes the merger rate per unit volume per unit number density of the two progenitors. Simple models suggest that the kernel likely increases with mass (Appendix~\ref{app:kernel_models}). We use the sum kernel to capture the positive mass dependence:
\begin{equation}
    K(M,m,t) = B(t)(M+m)\,. \label{eq:sum_kernel}
\end{equation}
This kernel form is mathematically well-understood, with available series solutions in general and closed-form ones for some specific initial distribution (e.g., \cite{Scott1968}). To connect with other theoretical works, we note that the specific BH major merger rate (with dimensions of inverse time) is approximately given by $K(M,M)\Phi(M)$. The local galaxy stellar mass function $\Phi(M_*)$ is a weakly decreasing function proportional to $M_*^{\alpha_1}$ for $10^{9}~\msun\lesssim M_*\lesssim 10^{11}~\msun$ ($\alpha_1=-0.35 \pm 0.18$ in Ref.~\cite{Baldry2012}). Thus, the sum kernel predicts that the specific galaxy major merger rate in this mass range scales as $M_*^{\alpha_1+1}$, which is in approximate agreement with semi-empirical, simulation-based, and semi-analytical studies (\cite{Hopkins2010a,RodriguezGomez2015,OLeary2021,Husko2022}, although the rates for $M_*>10^{11}~M_\odot$ suggest a stronger-than-linear $M_*$-dependence for the kernel). We assume the same kernel for all three types of mergers. 

With the sum kernel, integration of Equation~(\ref{eq:coagulation_h}) over mass gives the evolution of the heavy-seed BH number density:
\begin{equation}
    \frac{dn_h}{dt} = -\int_0^\infty\psi_{hh}(M,t)d\log M = -B(t)\rho_h(t)n_h\,,\label{eq:number_density_integral}
\end{equation}
where $\rho_h$ is the heavy-seed BH mass density defined analogously as in Equation~(\ref{eq:mass_density_conservation}). We rewrite Equation~(\ref{eq:number_density_integral}) as
\begin{equation}
    \frac{dn_h}{dt} = -\frac{n_h}{\tau_m}\,,\label{eq:number_decay}
\end{equation}
where $\tau_m$ characterizes the BH number density decay; note that this timescale does not correspond to an individual merger process but rather represents a statistical property of the system. A short timescale indicates frequent mergers. We assume this parameter remains constant over time, leading to an exponential decay in the number density, as described by Equation~(\ref{eq:number_decay}). Over time, $\rho_h$ will increase due to accretion and heavy-light mergers, so a constant $\tau_m$ implies a reducing $B(t)$. This deceleration qualitatively agrees with the behavior of dark-matter halo mergers \cite{Fakhouri2010}. As for the number density of the light-seed BHs, we similarly integrate Equation~(\ref{eq:coagulation_l}) to obtain
\begin{equation}
    \frac{dn_l}{dt} = -\frac{n_l}{\tau_m}\left[1+\frac{\rho_l}{\rho_h}\left(1+\frac{n_h}{n_l}\right)\right]\,.\label{eq:number_decay_l}
\end{equation}
The light-seed BH number density thus decays faster than the heavy-seed counterpart.

\subsection{MCMC-fitting to observed LF at $z\sim5$}
\label{subsec:MCMC}
\begin{table*}[ht]
    \centering
    \begin{tabular}{ccccccccc}
        \toprule
        Symbol & $n_{0h}$ & $n_{0l}$ & $\tau_m$ & $\tau$ & $\delta$ & $\lambda_{0h}$ & $\lambda_{0h}/\lambda_{0l}$ & $\alpha$  \\\midrule
        Unit & $\rm Mpc^{-3}$ & $\rm Mpc^{-3}$ & Myr & Myr & - & - & - & - \\
        Definition & Sect.~\ref{subsec:seeding} & Sect.~\ref{subsec:seeding} & Eq.~(\ref{eq:number_decay}) & Sect.~\ref{subsec:accretion} & Eq.~(\ref{eq:delta}) & Eq.~(\ref{eq:erdf}) & Sect.~\ref{subsec:accretion} & Eq.~(\ref{eq:erdf}) \\
        Vary in MCMC? & &&&&& \\
        -- heavy-only & $\times$ & N/A & $\times$ & \checkmark & \checkmark & \checkmark & N/A & \checkmark \\
        -- heavy+light & $\times$ & $\times$ & \checkmark & \checkmark & \checkmark & \checkmark & \checkmark & \checkmark \\
        \bottomrule
    \end{tabular}
    \caption{Summary of the model parameters. The \checkmark ($\times$) symbol means the parameter is varied (fixed) in each MCMC run. }
    \label{tab:free_params}
\end{table*}

The free parameters in this work have been described in the above sections and are summarized in Table~\ref{tab:free_params}. For a given set of parameters, we evolve the BHMF from $z=20$ through $z=5$ combining the accretion and merger treatments. The cosmic time interval is split into durations of equal length $\tau$ (any remainder forms an additional duration). We first advance one accretion time step, updating the BHMF using the analytical formula in Equation~(\ref{eq:BHMF_evolution_accretion}). Then, to account for mergers, we split the same accretion time step into smaller steps and update the accreted BHMF using Equation~(\ref{eq:coagulation}). This is done numerically, with details described in Appendix~\ref{app:numerical_mergers}. The accretion--merger calculation is repeated for the next duration of $\tau$ until the terminal redshift.

Following Ref.~\cite{Li2023}, we convolve the model BHMF at $z=5$ with the ERDF in Equation~(\ref{eq:erdf}) to obtain the intrinsic AGN bolometric LF, which is then converted to the unobscured LF in the rest-frame 1450~\AA\ band:
\begin{equation}
    \Phi_{M_{1450}} = (1-f_{\rm obsc})\int_0^\infty\frac{d\log L_{\rm bol}}{dM_{1450}} g(\tilde{\lambda}) \Phi_{M_\bullet}d\log M_\bullet\,, \label{eq:LF}
\end{equation}
where $M_{1450}$ is the rest-frame ultraviolet (UV) absolute magnitude at 1450~\AA\ and $\tilde{\lambda}\equiv L_{\rm bol}(M_{1450}) / L_{\rm Edd}(M_\bullet)$. The relation $L_{\rm bol}(M_{1450})$ is calculated using a bolometric correction $f_{1450}^{\rm bol}=4.4$ (read from Figure 12 in Ref.~\cite{Richards2006}), and the luminosity-dependent obscuration factor $f_{\rm obsc}$ is adopted from X-ray AGN observations at $z<5$ (\cite{Ueda2014}, their Equation 3).

To compare the model with data, we use the rest-UV-selected unobscured quasar LF from Ref.~\cite{Niida2020}, which covers the $1450~$\AA\ absolute magnitudes $-30~{\rm mag}<M_{1450}<-26$~mag based on the Sloan Digital Sky Survey and $-26~{\rm mag}<M_{1450}<-22$~mag observed with the Subaru Hyper Suprime-Cam. Additionally, we incorporate unobscured AGN candidates selected with JWST photometry \cite{Guo24}, which cover the UV magnitudes $-21~{\rm mag}<M_{1450}<-16$~mag. We note that the latter data set still awaits spectroscopic confirmation and gives an upper limit of the AGN abundance within the magnitude range. However, the abundance estimate agrees with those based on spectroscopically confirmed unobscured AGN samples \cite{Harikane2023,Maiolino2023}, though the statistics are limited. We exclude the data fainter than $-17.5$~mag due to potential incompleteness.

In our first set of fittings with only heavy-seed BHs, we manually vary $n_{0h}$ and $\tau_m$ to understand their effects on the model outputs and let MCMC explore the rest of the parameter space. In our second set of fittings including heavy- and light-seed BHs, $n_{0h}$ and $n_{0l}$ are controlled while $\tau_m$ and $\lambda_{0h}/\lambda_{0l}$ vary in the MCMC runs. We set up the same prior as in Ref.~\cite{Li2023} for $\tau,\delta,\lambda_{0h}$ (our $\lambda_{0h}$ corresponds to $\lambda_0$ in their work), and $\alpha$, and use a $\chi^2$ likelihood. Where needed, we use a log-uniform prior for $10^2~{\rm Myr}\leq\tau_m\leq10^6~{\rm Myr}$, as shorter merger timescales produce an unrealistically small number of $z\sim5$ quasars while mergers become negligible for longer $\tau_m$. We introduce a uniform prior for $1\leq\lambda_{0h}/\lambda_{0l}\leq3$; the light-seed BHs would have difficulty reproducing JWST AGNs if $\lambda_{0l}$ were too low. The parameter variation is summarized in Table~\ref{tab:free_params}.

We use the \texttt{emcee} MCMC Python package \cite{ForemanMackey2013}. In each run, 128 walkers are sampled for 3000 steps. The posterior distribution is tested to converge using doubled step numbers. 

\subsection{Observed GW event rates from mergers}
\label{subsec:GW_method}
After the best-fit parameter values are determined, the merger process in the model implies a differential merger event rate in the observer's frame, $\mathcal{R}_{ij}(M_\bullet,z)$ ($i,j=h,l$), or the all-sky event number per unit observer's year per unit logarithmic remnant BH mass per unit redshift. It is given by
\begin{equation}
    \mathcal{R}_{ij}(M,z) = \psi_{ij}(M,t(z))\frac{dV_C}{dz}\frac{1}{1+z}\,, \label{eq:observed_rate}
\end{equation}
where $dV_C/dz$ is the differential comoving volume. 

According to Equations~(\ref{eq:number_density_integral}), (\ref{eq:number_decay}), and (\ref{eq:observed_rate}), the observed heavy-heavy merger rate $N_{hh}(z)$ earlier than a given redshift $z$ only depends on $n_{0h}$ and $\tau_m$ but not on the BHMF profiles of the two populations:
\begin{gather}
    N_{hh}(z) = \int_z^{z_0}\int_0^\infty \mathcal{R}_{hh}(M,z')d\log Mdz' \nonumber \\
    = \frac{n_{0h}}{\tau_{\rm m}}\int_z^{z_0}\exp\left(-\frac{t(z')-t_0}{\tau_{\rm m}}\right)\frac{{\rm d}V_C}{{\rm d}z'}\frac{1}{1+z'}{\rm d}z'\,, \label{eq:total_merger}
\end{gather}
where $z_0=20$ and $t_0\equiv t(z_0)=179~\rm Myr$. In the limit of $\tau_{\rm m}\gg t(z)$, the merger rate decreases as $N_{hh}(z)\propto \tau_{\rm m}^{-1}$. On the other hand, the merger rate saturates in the limit of shorter merger timescales:
\begin{align}
    N_{hh}^{\rm max} & \equiv \lim_{\tau_{\rm m}\rightarrow 0}N_{hh} = \frac{n_{0h}}{1+z_0}\left(\frac{dV_C}{dz}\left|\frac{dz}{dt}\right|\right)_{z_0} \nonumber \\
    &= 0.46\left(\frac{n_{0h}}{10^{-3}~\rm Mpc^{-3}}\right)~\rm yr^{-1}\,. \label{eq:rate_upper_limit}
\end{align}
This sets an upper bound for the observed heavy-heavy merger event rate, which will be reached if the number density promptly vanishes, i.e., all BHs merge right after birth. Importantly, this bound applies to any merger model with a well-defined initial number density. The generality of this limit (e.g., its independence of the coagulation formalism and the exponential number density decay) is demonstrated in Appendix~\ref{app:upper_limit}. The total merger rate is also constrained by this limit, with $N_{hh}$ replaced with $N_{h+l}\equiv N_{hh}+N_{hl}+N_{lh}+N_{ll}$ and $n_{0h}$ replaced with $n_{0h}+n_{0l}$;
namely, 
\begin{align}
    N_{h+l}^{\rm max} & \equiv \lim_{\tau_{\rm m}\rightarrow 0}N_{h+l} 
    = \frac{n_{0h}+n_{0l}}{1+z_0}\left(\frac{dV_C}{dz}\left|\frac{dz}{dt}\right|\right)_{z_0} 
    \nonumber \\
    &= 4.6\times 10^2 \left(\frac{n_{0h}+n_{0l}}{1~\rm Mpc^{-3}}\right)~\rm yr^{-1}\,. \label{eq:rate_upper_limit_hl}
\end{align}

Further, we evaluate the detectability of merger events by comparing the signal spectrum and the detector noise curves. We generate GW waveforms using PhenomD \cite{Husa2016,Khan2016}, a phenomenological model for spin-aligned binary BHs, and its Python implementation, \texttt{BOWIE} \cite{Katz2019}. We assume for simplicity that all merging MBHs have no spin, so the observed GW will only depend on the remnant mass $M_\bullet$, the secondary-to-primary mass ratio $q\leq1$, and redshift $z$. On the detector side, we mainly consider the noise curves of LISA \cite{Babak2021} and B-DECIGO \cite{Kawamura2023}. We take a nominal signal-to-noise (SNR) threshold of $\rm SNR_{thresh}=8$, above which a given event is assumed to be detectable. 

\begin{figure}
   \centering
   \includegraphics[width=0.47\textwidth]{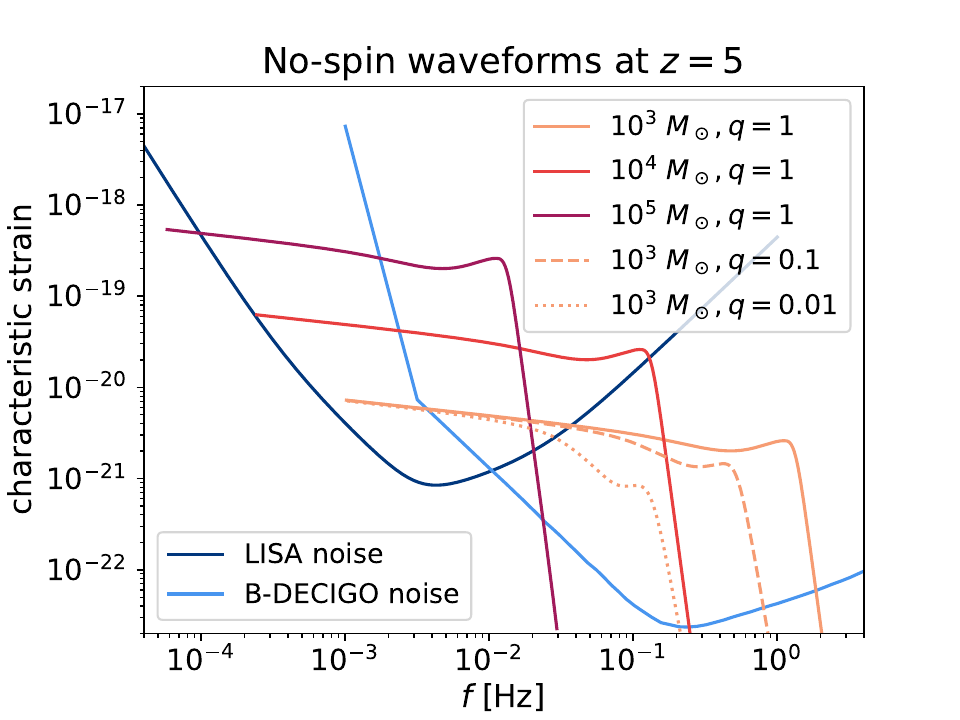}
   \caption{GW waveforms of no-spin BH mergers compared with LISA and B-DECIGO noise amplitudes. The horizontal axis denotes the 
   observed frequency. The legend denotes the chirp mass $\mathcal{M}$ and the secondary-to-primary mass ratio $q$. }
   \label{fig:GW_hc}
\end{figure}

Figure~\ref{fig:GW_hc} shows the strain amplitudes of merger events of different binary black hole chirp masses and mass ratios. GW waveforms are calculated from the inspiral phase one year before the merger through the ringdown phase (using a longer detected inspiral time would slightly extend the low-frequency end of the waveforms). The signal curves with $q=1$ are copies of one profile shifted to different frequencies and amplitudes. Different mass ratios with the same chirp mass have identical inspiral spectra, but the merger--ringdown amplitude is significantly suppressed for unequal-mass mergers. Therefore, major mergers are likely the most readily detected gravitational wave sources if other factors (e.g., event rates) are controlled. 

Figure~\ref{fig:GW_hc} also presents the noise curves of LISA and B-DECIGO\footnote{Ref.~\cite{Kawamura2023} presented the B-DECIGO noise curve for $f>3\times10^{-3}~\rm Hz$. We assume a pessimistic dimensionless noise amplitude $\propto f^{-6}$ for lower frequencies (cf. \cite{Katz2019}).}. The LISA devices will follow a heliocentric orbit while the B-DECIGO ones will circle the Earth with shorter interferometer arm lengths. As a result of the different designs, B-DECIGO has lower noises for $f\gtrsim10^{-2}~\rm Hz$ (thus more suitable for relatively light MBHs), but LISA performs better at lower frequencies.

\section{Results: heavy-seed BHs}
\label{sec:heavy}

\begin{table*}[ht]
    \centering
    \begin{tabular}{cccccccc}
         \toprule
         $\log n_{0h}$ & $\log\tau_m$ &  $\tau$ & $\log\delta$ & $\lambda_{0h}$ & $\alpha$ & $\overline{\lambda_{0h}}$ & $\ln P_{\rm post}$\\
         \small{(Mpc$^{-3}$)} & \small{(Myr)} & \small{(Myr)} & - & - & - & - & - \\\midrule
         -3 & 3 & 28.1 & -1.474 & 0.324 & 0.623 & 0.230 & -293 \\
         & & ($27^{+9}_{-7}$) & ($-1.54^{+0.18}_{-0.28}$) & ($0.33^{+0.09}_{-0.07}$) & ($0.62^{+0.21}_{-0.18}$) & & \vspace{4pt}\\
         -3 & 4 & 34.1 & -2.08 & 0.274 & 0.873 & 0.253 & -194 \\
         & & ($33^{+9}_{-8}$) & ($-2.3^{+0.4}_{-0.5}$) & ($0.28^{+0.07}_{-0.05}$) & ($0.84^{+0.21}_{-0.20}$) & & \vspace{4pt}\\
         -3 & 5 & 34.1 & -2.09 & 0.274 & 0.880 & 0.255 & -184 \\
         & & ($33^{+9}_{-7}$) & ($-2.3^{+0.4}_{-0.5}$) & ($0.28^{+0.07}_{-0.05}$) & ($0.85^{+0.21}_{-0.20}$) & & \vspace{6pt}\\
         -2 & 3 & 57.4 & -1.421 & 0.307 & 0.358 & 0.162 & -191 \\
         & & ($55^{+17}_{-14}$) & ($-1.50^{+0.22}_{-0.44}$) & ($0.31^{+0.12}_{-0.07}$) & ($0.36^{+0.21}_{-0.22}$) & & \vspace{4pt}\\
         -2 & 4 & 57.6 & -1.76 &  0.279 & 0.533 & 0.182 & -123 \\
         & & ($55^{+17}_{-13}$) & ($-2.0^{+0.4}_{-0.7}$) & ($0.29^{+0.09}_{-0.07}$) & ($0.52^{+0.23}_{-0.21}$) & & \vspace{4pt}\\
         -2 & 5 & 57.5 & -1.80 & 0.274 & 0.557 & 0.184 & -118 \\
         & & ($53^{+19}_{-14}$) &  ($-2.0^{+0.4}_{-0.7}$) & ($0.29^{+0.11}_{-0.07}$) & ($0.51^{+0.24}_{-0.23}$) & & \\
         \bottomrule
    \end{tabular}
    \caption{MCMC outputs in the heavy-seed-only runs. The odd rows give the best-fit values for the four accretion-controlling parameters and the even rows give the median values with upper and lower uncertainties in brackets. }
    \label{tab:best_fit_heavy}
\end{table*}

\begin{figure*}[ht]
        \centering
        \includegraphics[width=0.98\textwidth]{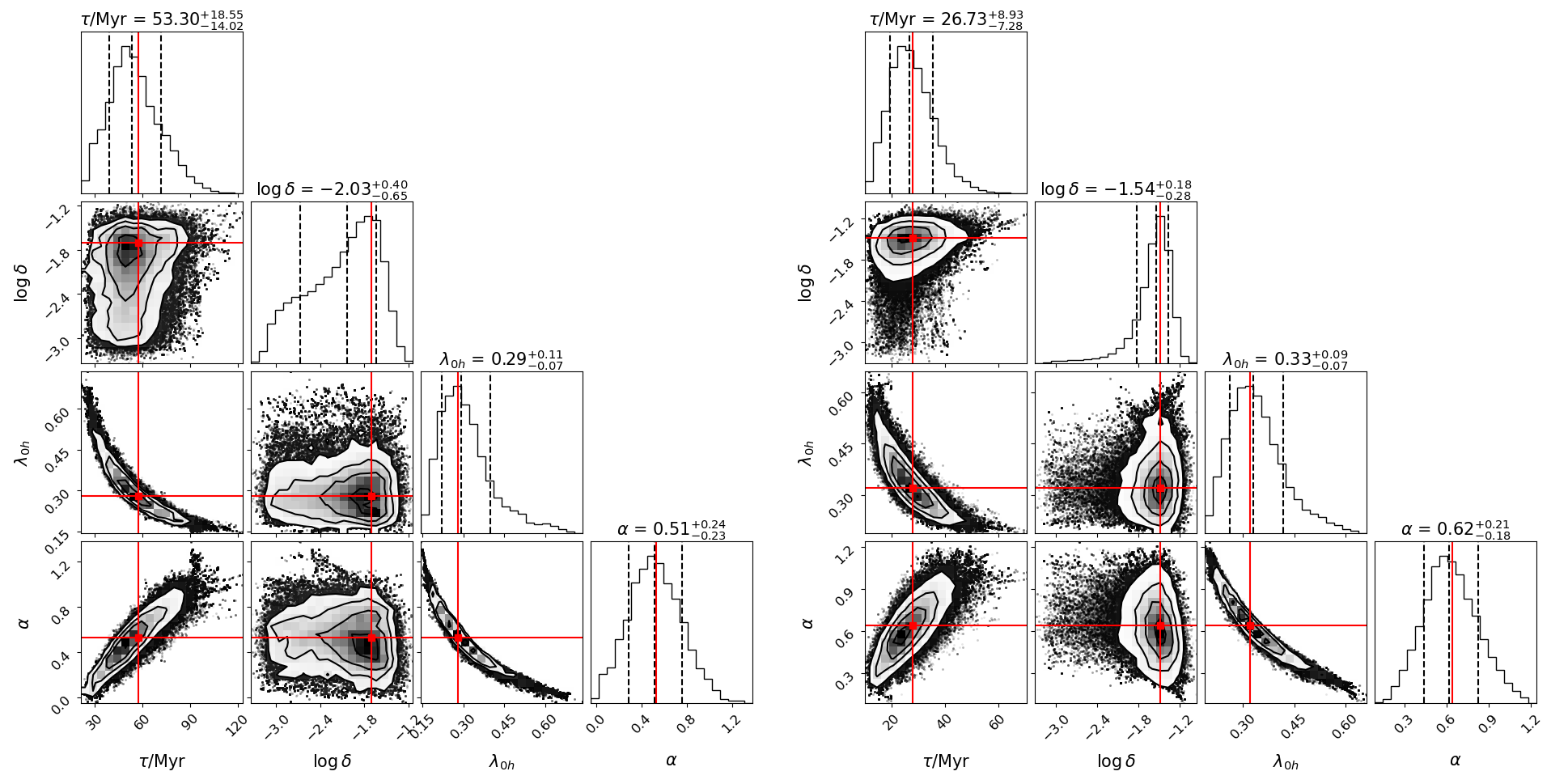}

    \caption{MCMC corner plots corresponding to the heavy-seed-only run with $(\log n_{0h}/{\rm Mpc^{-3}}, \log\tau_m/{\rm Myr})=(-2,5)$ (\textit{left}) and $(-3,3)$ (\textit{right}). The diagonal panels present the posterior distribution of each parameter, with the medium and 1-$\sigma$  marked with dashed lines and annotated atop. The off-diagonal panels show the two-parameter correlation. Red lines and points denote the location of the best-fit value.}
    \label{fig:MCMC_corner_heavy}
\end{figure*}

\begin{table*}[ht]
    \centering
    \begin{tabular}{ccccccc}
         \toprule
         $\log n_{0h}$ & $\log\tau_m$ & $n_{h,z5}$ & $\rho_{h,z5}$ & $N_{hh}$ & $N_{hh,q>0.10}$ & $N_{hh,q>0.25}$\\
         \small{(Mpc$^{-3}$)} & \small{(Myr)} & \small{(Mpc$^{-3}$)} & \small{($M_\odot$\,Mpc$^{-3}$)} & \small{(yr$^{-1}$)} & \small{(yr$^{-1}$)} & \small{(yr$^{-1}$)} \\\midrule
         -3 & 3 & $3.7\times10^{-4}$ & $2.4\times10^{3}$ & $2.1\times10^{-1}$ & $9.6\times10^{-2}$ & $6.0\times10^{-2}$ \\
         -3 & 4 & $9.1\times10^{-4}$ & $3.4\times10^3$ & $3.1\times10^{-2}$ & $1.5\times10^{-2}$ & $9.1\times10^{-3}$ \\
         -3 & 5 & $9.9\times10^{-4}$ & $3.5\times10^3$ & $3.2\times10^{-3}$ & $1.5\times10^{-3}$ & $9.6\times10^{-4}$ \\
         -2 & 3 & $3.7\times10^{-3}$ & $5.9\times10^3$ & $2.1\times10^{0}$ & $8.2\times10^{-1}$ & $5.0\times10^{-1}$\\
         -2 & 4 & $9.1\times10^{-3}$ & $8.7\times10^3$ & $3.1\times10^{-1}$ & $1.3\times10^{-1}$ & $7.9\times10^{-2}$ \\
         -2 & 5 & $9.9\times10^{-3}$ & $9.1\times10^3$ & $3.2\times10^{-2}$ & $1.4\times10^{-2}$ & $8.4\times10^{-3}$ \\
         \bottomrule
    \end{tabular}
    \caption{BHMF and merger properties in the heavy-seed-only models, calculated using the best-fit parameters in each run.}
    \label{tab:merger_rate_heavy}
\end{table*}

\begin{figure}[ht]
    \centering
    \includegraphics[width=0.5\textwidth]{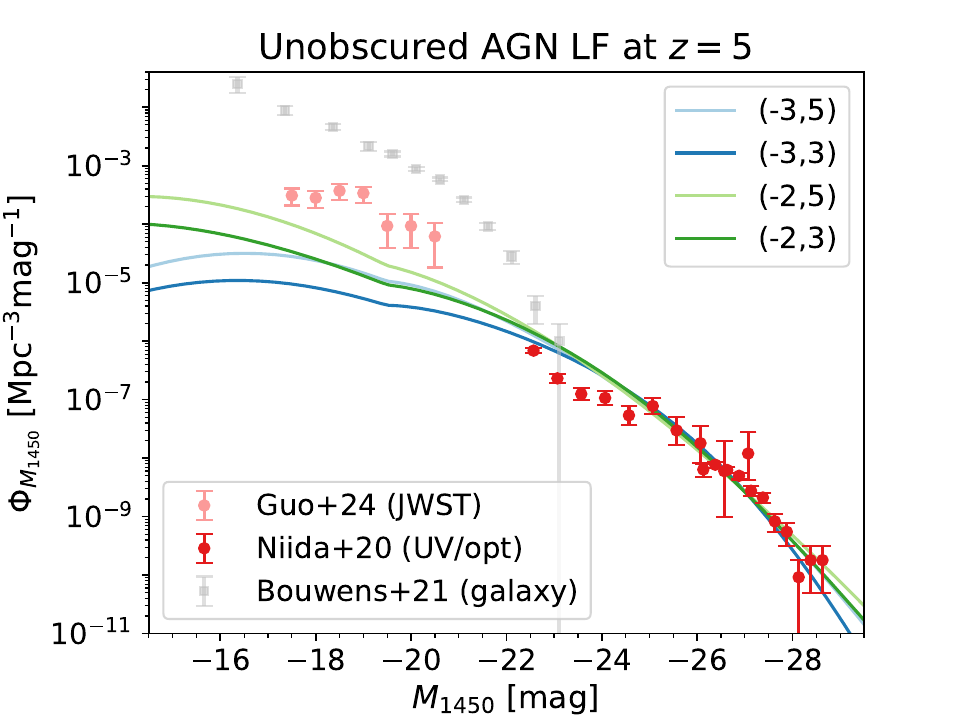}
    \caption{Model and observed unobscured AGN LF at $z=5$ (heavy-seed only). The legends for the model curves indicate $(\log n_{0h}/{\rm Mpc^{-3}}, \log\tau_m/{\rm Myr})$. Data are taken from ground-based, rest-UV/optical-selected AGNs (red \cite{Niida2020}) and JWST unobscured AGN candidates (pink \cite{Guo24}). The UV galaxy LF \cite{Bouwens2021} is also plotted (grey) for comparison.}
    \label{fig:QLF_heavy}
\end{figure}

\begin{figure}[ht]
    \centering
    \includegraphics[width=0.5\textwidth]{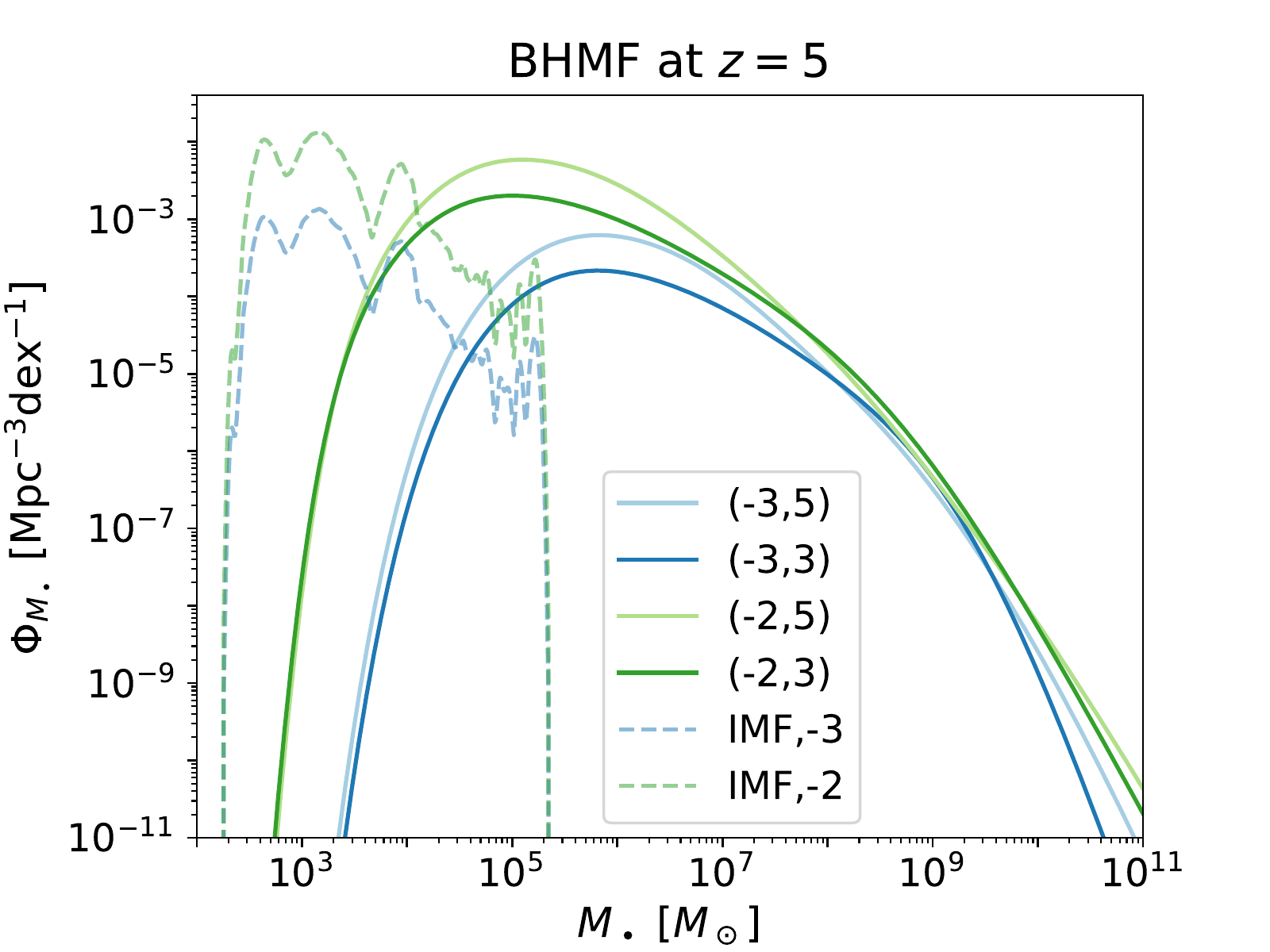}
    \caption{Model BHMF (heavy-seed only, including the unobscured and obscured population). The legends for the model curves at $z=5$ (solid) indicate $(\log n_{0h}/{\rm Mpc^{-3}}, \log\tau_m/{\rm Myr})$. The seed MFs at $z=20$ in the mass range $M_\bullet>10^4~M_\odot$ are plotted in dashed curves for comparison. ``IMF, -3'' stands for the IMF with $n_{0h}=10^{-3}~\rm Mpc^{-3}$ and ``-2'' for $10^{-2}~\rm Mpc^{-3}$.}
    \label{fig:BHMF_heavy}
\end{figure}

In the heavy-seed-only model, we perform six MCMC runs with different $n_{0h}$ and $\tau_m$ values. The default in Ref.~\cite{Li2023} is\footnote{The exact value is $9.9\times10^{-4}~\rm Mpc^{-3}$.} $n_{0h}=10^{-3}~\rm Mpc^{-3}$, a prediction based on the number density of dark-matter halos with masses $M_{\rm halo}\geq10^{11}~\msun$ at $z=6$; we additionally explore ten times this value (i.e., $n_{h0}=10^{-2}~\rm Mpc^{-3}$) to see if the heavy-seed BH population alone may explain the abundance of JWST AGNs. After some initial tests, we choose the $\tau_m$ values as $10^3$, $10^4$, and $10^5~{\rm Myr}$. The shortest $\tau_m$ is approximately equal to the total elapsed time of $t_{z5}-t_{z20}=995~\rm Myr$. The results for the six cases are listed in Table~\ref{tab:best_fit_heavy}.

We first discuss the parameter posterior distribution in the run with $n_{0h}=10^{-2}~\rm Mpc^{-3}$ and $\tau_m=10^5~\rm Myr$, visualized in the left panel in Figure~\ref{fig:MCMC_corner_heavy}. The histograms along the diagonal show that the parameters $\tau$, $\lambda_{0h}$, and $\alpha$ are relatively tightly constrained. In contrast, the distribution of $\log\delta$ is wide and has a peak on the right of the median,
indicating that suppression of high-mass BH growth is modestly favored. In the off-diagonal panels, one observes that $\tau$ and $\alpha$ are positively correlated with each other and anti-correlated with $\lambda_{0h}$. The anti-correlations, also present in Ref.~\cite{Li2023}, are likely because all the three parameters have similar effects of enhancing the abundance of the most luminous quasars. A long accretion timescale would enable sustained growth for MBHs with large Eddington ratios, thus boosting the high-mass end of the BHMF. A large characteristic Eddington ratio would increase the likelihood of rapid accretion. A large power-law index in the ERDF would increase the mean Eddington ratio. On the other hand, the $\tau$--$\alpha$ correlation was not obvious in Ref.~\cite{Li2023} and might not have a straightforward interpretation. The posterior plots in a different run of $n_{0h}=10^{-3}~\rm Mpc^{-3}$ and $\tau_m=10^3~\rm Myr$ (right panel in Figure~\ref{fig:MCMC_corner_heavy}) are qualitatively similar except for a sharper $\log\delta$ distribution, which implies a stronger model preference to retard the growth of MBHs of mass $\gtrsim10^8~M_\odot$.

Table~\ref{tab:best_fit_heavy} reports the best-fit parameter values and the maximum posterior probability in each run, along with the posterior distributions characterized by the median values with upper and lower uncertainties. The mean Eddington ratios $\overline{\lambda_{0h}}$ are also calculated using Equation~(\ref{eq:erdf}) and the best-fit $\lambda_{0h}$ and $\alpha$ values. Figures~\ref{fig:QLF_heavy} and \ref{fig:BHMF_heavy} present the corresponding best-fit unobscured AGN LF and the total (unobscured and obscured) BHMF. For clarity, these figures only present the curves with $\tau_m=10^5$ or $10^3~\rm Myr$; the $\tau_m=10^4~\rm Myr$ cases are very similar to the $10^5~\rm Myr$ ones. The first four columns of Table~\ref{tab:merger_rate_heavy} list the total $z=5$ number and mass density of the heavy-seed BHs, $n_{h,z5}$ and $\rho_{h,z5}$. Note that $n_{h,z5}$ is calibrated with analytical calculations (Equation~\ref{eq:number_decay}; see also Appendix~\ref{app:numerical_mergers}) while $\rho_{h,z5}$ is numerically measured from the BHMF.

The number density of seed BHs significantly influences the parameter distribution. A higher seed abundance results in a decreased ERDF power-law index, which gives a broader range of Eddington ratios. The additional BHs thus significantly raise the faint end of the AGN LF,  bringing it closer to the observed abundance and explaining why higher values of $n_{0h}$ tend to yield higher posterior probabilities. On the other hand, the lower $\alpha$ reduces the mean Eddington ratio, which necessitates higher BH masses at $z\sim5$ at the massive end of the BHMF to reproduce the quasar luminosities: see, e.g., the (-3, 5) v.s. (-2, 5) curve in Figure~\ref{fig:BHMF_heavy}. This is realized with longer best-fit accretion timescales, although the value is still on the order of several $10~\rm Myr$.

The case with $n_{0h}=10^{-3}~{\rm Mpc^{-3}}$ and $\tau_m=10^5~\rm Myr$ is similar to the $f_{\rm seed}=1$ model in Ref.~\cite{Li2023}\footnote{Their model BHMF and LF are publicly available on GitHub: \url{https://github.com/WenxiuLiii/BHMF_QLF_data}}. Their study found larger best-fit $\lambda_{0h}\ (=0.96)$ and smaller $\alpha\ (=-0.06)$, leading to a broader distribution in both the LF and BHMF. The difference arises from the fitting data adopted. This earlier work was calibrated to ground-based unobscured LF and observed BHMF at $z\sim6$, where the bright-end ($M_{1450}<-27~\rm mag$) LF cutoff was loosely bound. Furthermore, the new JWST data requires abundant MBHs to have moderate luminosity. As a result, the $z\sim5$ LF in our work favors AGNs concentrated within the interval $-27~\lesssim M_{1450}\lesssim -17~{\rm mag}$, with fewer brighter AGNs. The mean Eddington ratio we derive ($\overline{\lambda_{0h}}=0.26$) is comparable to the one found in Ref.~\cite{Li2023} ($\overline{\lambda_{0h}}=0.21$).

Back to this work, the influence of the merger timescale depends on its value. One may expect that mergers have negligible effects on the MBH evolutionary process once $\tau_m\gg t_{z5}-t_{z20}=995~\rm Myr$. Indeed, $\tau_m=10^4~\rm Myr$ and $10^5~\rm Myr$ do not strongly influence the total MBH number density and give similar posterior parameter distributions. Therefore, they may be regarded as convergent to an accretion-dominant scenario. On the other hand, in the case of $\tau_m=10^3~\rm Myr$, the significant number density reduction indicates that mergers are becoming important: the parameter distributions deviate from the accretion-dominant counterparts. In both cases of $n_{0h}=10^{-3}$ and $10^{-2}~\rm Mpc^{-3}$, shorter merger timescales prefer stronger suppression of the growth of the heaviest MBHs (i.e., larger $\delta$) and more inactive MBHs (i.e., smaller $\alpha$). The average Eddington ratio is reduced despite mildly larger $\lambda_{0h}$. These illustrate the complementary role of accretion and mergers: the sum kernel of the merger process encourages rapid assembly of the most massive population, so accretion must be weakened to match observation.  

The rarity of heavy seeds in our model has two implications. Firstly, even the best-fit LF curves fall short of the JWST AGN candidate abundance by a factor of several or more. This indicates that the heavy-seed population alone has difficulty explaining the data, and suggests that additional MBHs may contribute to the LF. Secondly, assuming that heavy-seed BHs account for the LF at $M_{1450}\lesssim-17~\rm mag$, the fitting results impose strict limits on mergers. Table~\ref{tab:merger_rate_heavy} lists the total observed event rates as well as the rates for mergers with $q>0.10$ and $q>0.25$, respectively. We define $q>0.25$ as major mergers and $0.10<q\leq0.25$ as minor ones. The merger rates of heavy-heavy seed BHs, $N_{hh}$, agree well with the analytical predictions in Equation~(\ref{eq:total_merger}), which is $>2$ times lower than the upper bound given in Equation~(\ref{eq:rate_upper_limit}) due to the large $\tau_m$.  Shorter $\tau_{\rm m}$ would struggle even more to sustain a sufficient number density of AGNs at the faint end, which is why we exclude cases with $\tau_{\rm m} < 10^3~{\rm Myr}$ from consideration. Importantly, models with the fiducial value of $n_{0h}=10^{-3}~{\rm Mpc}^{-3}$ have $N_{hh,q>0.25}<0.1~\rm yr^{-1}$; 
even with the most optimistic parameter set ($n_{0h}=10^{-2}~{\rm Mpc}^{-3}$ and $\tau_{\rm m}=10^3~{\rm Myr}$), the major merger rates remain below $1~\rm yr^{-1}$. In summary, while heavy-seed BHs can reproduce quasars at $z\sim5$, their rarity likely suggests an insufficient number density to explain fainter AGNs and a pessimistic merger rate for future GW observation.

\section{Results: heavy- and light-seed BHs}
\label{sec:hl}
\subsection{AGN LF and BHMF}

\begin{table*}[ht]
    \centering
    \begin{tabular}{cccccccccc}
         \toprule
         $\log n_{0h}$ & $\log n_{0l}$ &  $\tau$ & $\log\delta$ & $\lambda_{0h}$ & $\alpha$ & $\log\tau_m$ & $\lambda_{0h}/\lambda_{0l}$ & $\overline{\lambda_{0h}}$ & $\ln P_{\rm post}$\\
         \small{(Mpc$^{-3}$)} & \small{(Mpc$^{-3}$)} & \small{(Myr)} & - & - & - & \small{(Myr)} & - & - & - \\\midrule
         -3 & 0 & 19.5 & -2.98 &  0.926 & -0.027 &  5.59 & 1.1278 & 0.220 & -50.1 \\
         & & ($24^{+10}_{-6}$) & ($-2.7^{+0.5}_{-0.3}$) & ($0.74^{+0.25}_{-0.23}$) & ($0.06^{+0.19}_{-0.12}$) & ($5.5^{+0.4}_{-0.4}$) & ($1.128^{+0.023}_{-0.020}$) & \vspace{4pt}\\
         -2 & 0 & 36.8 & -2.76 &  0.566 & 0.037 &  4.34 & 1.0002 & 0.170 & -67.3 \\
         & & ($43^{+11}_{-8}$) & ($-2.3^{+0.5}_{-0.5}$) & ($0.47^{+0.12}_{-0.09}$) & ($0.12^{+0.13}_{-0.10}$) & ($4.5^{+0.8}_{-0.4}$) & ($1.010^{+0.014}_{-0.007}$) & \vspace{4pt}\\
         -3 & 1 & 35.4 & -2.39 & 0.774 & -0.058 & 5.329 & 1.404 & 0.182 & -46.9 \\
         & & ($35^{+11}_{-10}$) & ($-2.2^{+0.8}_{-0.6}$) & ($0.78^{+0.31}_{-0.20}$) & ($-0.08^{+0.14}_{-0.12}$) & ($5.36^{+0.36}_{-0.30}$) & ($1.40^{+0.06}_{-0.05}$) \vspace{4pt}\\
         -2 & 1 & 44.1 & -2.99 & 0.609 & -0.085 & 4.50 & 1.165 & 0.147 & -57.2 \\
         & & ($53^{+20}_{-12}$) & ($-2.6^{+0.5}_{-0.4}$) & ($0.48^{+0.16}_{-0.13}$) & ($0.03^{+0.17}_{-0.15}$) & ($4.8^{+0.5}_{-0.4}$) & ($1.15^{+0.05}_{-0.04}$) \vspace{4pt}\\
         \bottomrule
    \end{tabular}
    \caption{MCMC outputs in the heavy-and-light-seed runs. For parameters varied during the MCMC run, the odd rows give the best-fit values and the even rows give the median values with upper and lower uncertainties in brackets. }
    \label{tab:best_fit_hl}
\end{table*}

\begin{figure*}[ht]
    \centering
    \includegraphics[width=0.65\textwidth]{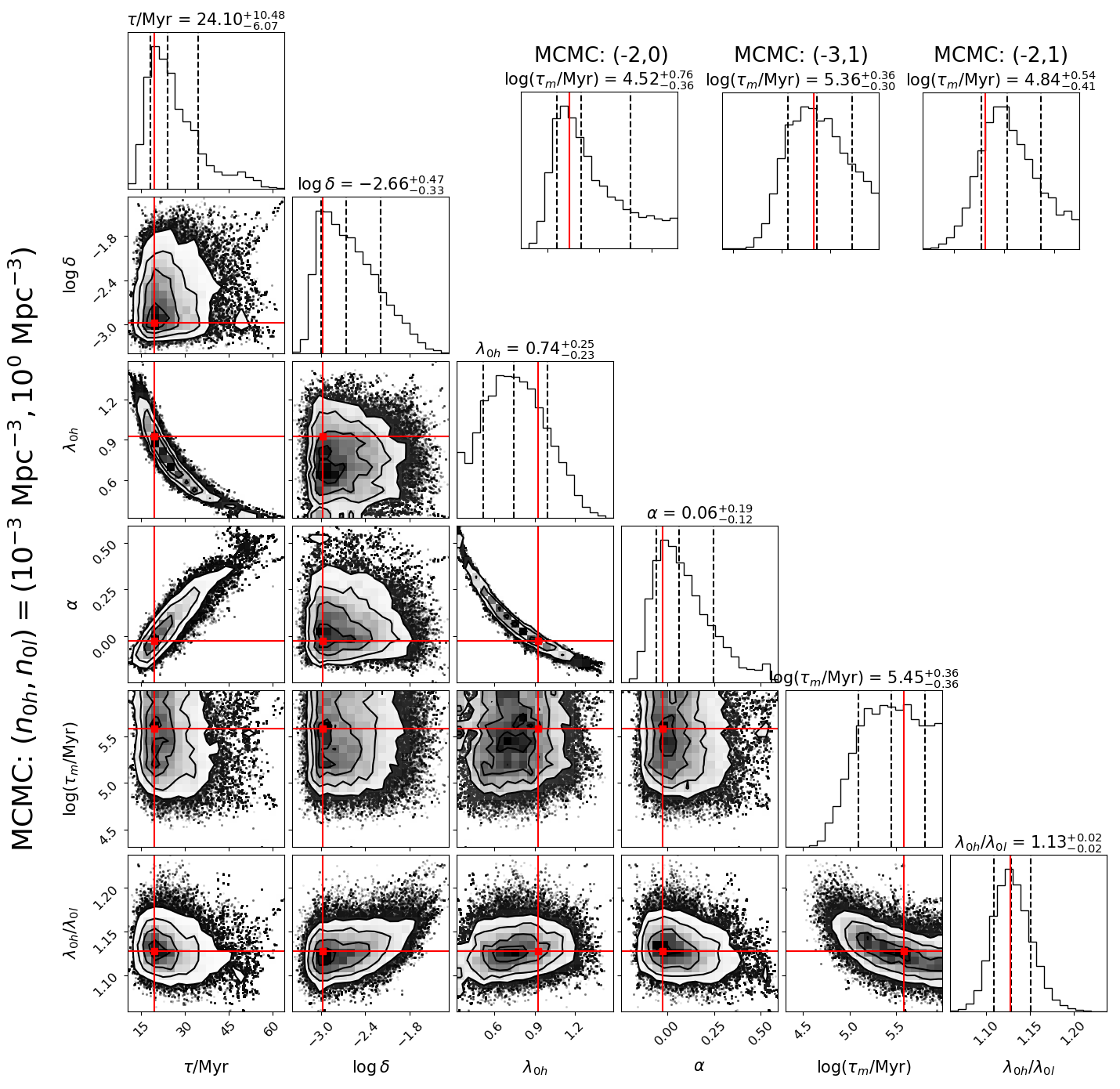}
    \caption{\textit{Bottom left, main}: MCMC corner plots corresponding to the heavy-and-light-seed run with $n_{0h}=10^{-3}~{\rm Mpc^{-3}}$ and $n_{0l}=10^0~{\rm Mpc^{-3}}$. Other details are the same as Figure~\ref{fig:MCMC_corner_heavy}. \textit{Top right}: posterior of $\log\tau_m$ in the three other runs.}
    \label{fig:MCMC_corner_hl}
\end{figure*}

\begin{table*}[ht]
    \centering
    \begin{tabular}{ccccccccc}
         \toprule
         $\log n_{0h}$ & $\log n_{0l}$ & $n_{h,z5}$ & $n_{l,z5}$ &  $\rho_{h,z5}$ & $\rho_{l,z5}$ & $N_{hh}$ & $N_{hl+lh}$ & $N_{ll}$\\
         \multicolumn{2}{c}{\small{(Mpc$^{-3}$)}} & \multicolumn{2}{c}{\small{(Mpc$^{-3}$)}} & \multicolumn{2}{c}{\small{($M_\odot$\,Mpc$^{-3}$)}} &  \small{(yr$^{-1}$)} & \small{(yr$^{-1}$)} & \small{(yr$^{-1}$)} \\\midrule
         -3 & 0 & $9.97\times10^{-4}$ & 0.913 & $1.86\times10^3$ & $4.46\times10^4$ & $8.42(2.20)\times10^{-4}$ & $8.34(0.382)\times10^{-1}$ & $3.15(2.00)\times10^1$ \\
         -2 & 0 & $9.56\times10^{-3}$ & 0.757 & $7.00\times10^3$ & $2.99\times10^4$ & $1.45(0.388)\times10^{-1}$ & $13.8(0.963)\times10^0$ & $7.19(4.03)\times10^1$ \\
         -3 & 1 & $9.95\times10^{-4}$ & 4.27 & $2.01\times10^3$ & $1.07\times10^5$ & $1.53(0.427)\times10^{-3}$ & $9.62(0.395)\times10^0$ & $2.48(1.60)\times10^3$ \\
         -2 & 1 & $9.69\times10^{-3}$ & 4.17 & $7.78\times10^3$ & $9.10\times10^4$ & $9.87(2.91)\times10^{-2}$ & $6.54(0.359)\times10^1$ & $2.15(1.30)\times10^3$\\
         \bottomrule
    \end{tabular}
    \caption{BHMF and merger properties in the heavy-and-light-seed models, calculated using the best-fit parameters in each run. Values in parentheses represent merger rates with $q>0.25$.}
    \label{tab:merger_rate_hl}
    \vspace{-5mm}
\end{table*}

\begin{figure}[t]
    \centering
    \includegraphics[width=0.5\textwidth]{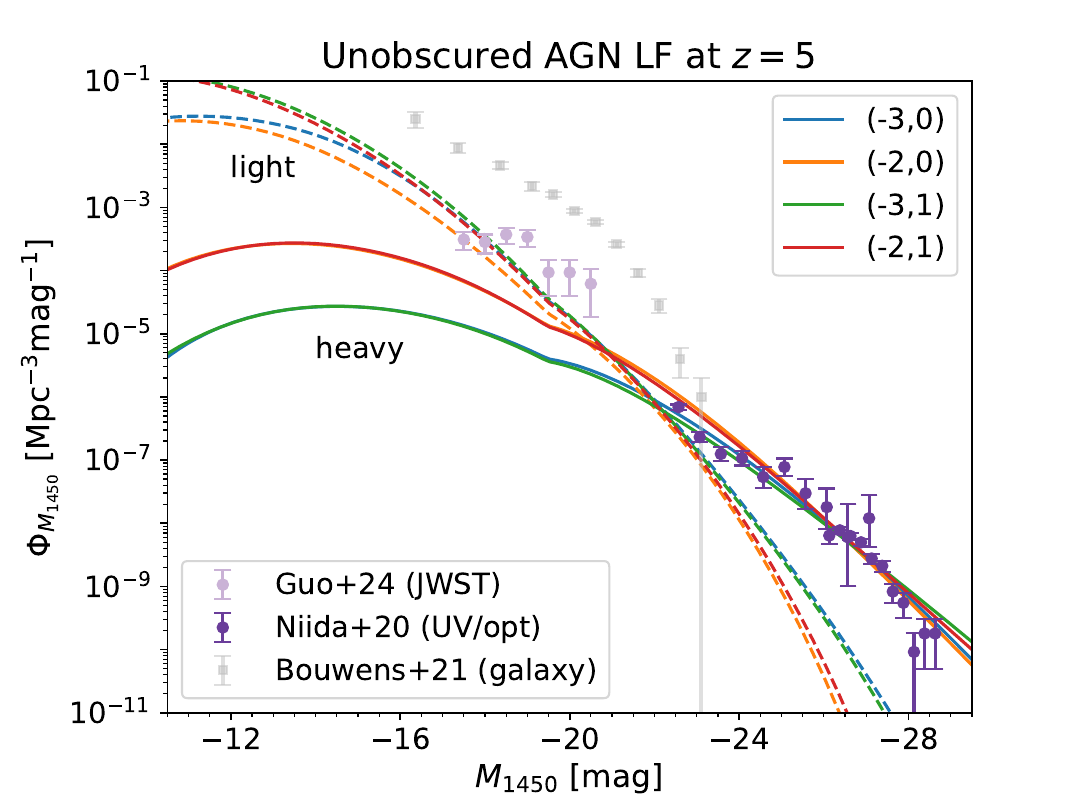}
    \caption{Model and observed unobscured AGN LF at $z=5$ (heavy-and-light-seed model). The legends for the model curves indicate $(\log n_{0h}/{\rm Mpc^{-3}},\log n_{0l}/{\rm Mpc^{-3}})$. Solid (dashed) curves represent the heavy- (light-) seed BH components. Data are taken from ground-based, rest-UV selected AGNs (dark purple \cite{Niida2020}) and JWST unobscured AGN candidates (pale purple \cite{Guo24}). The UV galaxy LF \cite{Bouwens2021} is also plotted (grey) for comparison. The orange and red solid lines mostly coincide.}
    \label{fig:QLF_hl}
\end{figure}

\begin{figure}[t]
    \centering
    \includegraphics[width=0.5\textwidth]{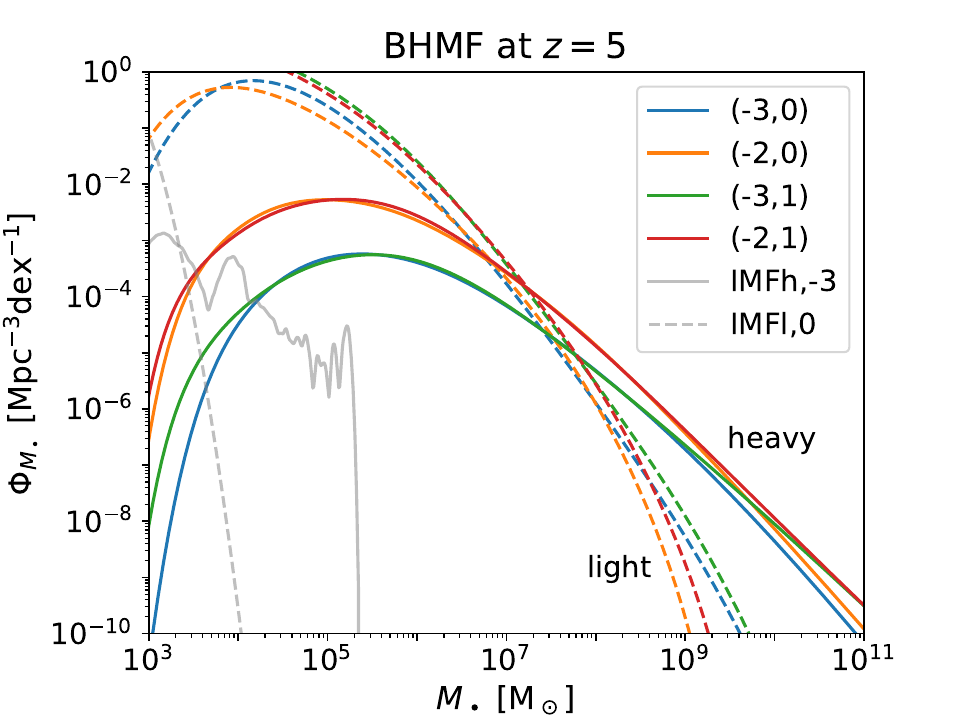}
    \caption{Model BHMF (heavy-and-light-seed, including the unobscured and obscured population). The legends for the model curves at $z=5$ (solid) indicate $(\log n_{0h}/{\rm Mpc^{-3}}, \log n_{0l}/{\rm Mpc^{-3}})$. Solid (dashed) curves represent the heavy- (light-) seed BH components. For comparison, ``IMFh, -3'' and ``IMFl, 0'' present the heavy- and light-seed IMFs with $n_{0h}=10^{-3}~\rm Mpc^{-3}$ and $n_{0l}=10^0~\rm Mpc^{-3}$.}
    \label{fig:BHMF_hl}
\end{figure}

In this section, we consider heavy-and-light-seed models. We perform four MCMC runs with different $n_{0h}$ and $n_{0l}$ values. In contrast to the previous section, we let MCMC find the best-fit $\tau_m$. The setup and outputs are summarized in Table~\ref{tab:best_fit_hl}. 

Figure~\ref{fig:MCMC_corner_hl} presents the posterior distribution of our fiducial case with $n_{0h}=10^{-3}~{\rm Mpc^{-3}}$ and $n_{0l}=10^{0}~{\rm Mpc^{-3}}$. The first four parameters are similarly distributed as those in the heavy-seed-only model in Section~\ref{sec:heavy} except that a smaller $\delta$ (i.e., approximately exponential growth) is now preferred. The merger timescale distribution here forms a broad peak at $\log(\tau_m/\rm Myr)=5.5$ and quickly declines at $<5.1$, disfavoring more frequent mergers. A mild decreasing trend on the right of the peak extends to the prior upper bound, $10^6~\rm Myr$, so fewer mergers than the best fit may adequately explain the data. The ratio $\lambda_{0h}/\lambda_{0l}$ forms a sharp peak at $1.13\pm0.02$. Thus, light- and heavy-seed BHs accrete with comparable typical rates (the latter are faster); even a slight deviation in $\lambda_{0h}/\lambda_{0l}$ significantly modifies the prediction because the difference is exponentially enlarged during the BH mass growth. The off-diagonal panels suggest that $\lambda_{0h}/\lambda_{0l}$ is weakly but positively correlated with $\log\delta$, and negatively with $\log\tau_m$. Increasing the value of $\delta$ implies that on average, less massive BHs accrete faster than their heavier counterparts at the same Eddington ratio, while decreasing the value of $\tau_m$ leads to rapid light-seed BH mergers (the heavy-seed BHs are not significantly influenced until $\tau_m\lesssim10^3~\rm Myr$). Both have similar effects as increasing $\lambda_{0l}$ relative to $\lambda_{0h}$ on the mass assembly of the light-seed BHs. 

The posterior profiles and correlation in other runs generally resemble the fiducial case. However, their distribution functions of $\log\tau_m$ decline faster on the right of the peak and thus indicate a tighter constraint on the merger rate, as shown on the top-right panels in Figure~\ref{fig:MCMC_corner_hl}. Additionally, in the run with $n_{0h}=10^{-2}~{\rm Mpc^{-3}}$ and $n_{0l}=10^0~{\rm Mpc^{-3}}$, the best-fit $\lambda_{0h}/\lambda_{0l}=1.0002$ almost reaches the lower bound of the prior, suggesting that allowing faster light-seed BH accretion in this case may better fit the observed AGN LF. 

Tables~\ref{tab:best_fit_hl} and \ref{tab:merger_rate_hl} show the MCMC outputs and BHMF properties of the four runs. The unobscured AGN LF and the total BHMF are visualized in Figures~\ref{fig:QLF_hl} and \ref{fig:BHMF_hl}. The substantially higher posterior probability indicates that introducing the light-seed population improves the fit. Indeed, the light-seed model LF curves reach abundances comparable to the JWST unobscured AGN candidates. Notably, in all four runs, the bright end of the LF, i.e., $M_{1450}\leq-22~\rm mag$, mainly comprises the heavy-seed population, whereas the light-seed population dominates the LF for $M_{1450}\geq-20~\rm mag$. The two types of MBHs explain the ground- and space-based AGNs respectively, in agreement with the conclusion in the previous section. 

How the best-fit values depend on the number density is similar to the previous section: higher $n_{0h}$ or $n_{0l}$ increases $\tau$ and decreases $\overline{\lambda_{0h}}$. The newly introduced parameter $\lambda_{0h}/\lambda_{0l}$ appears positively correlated with $n_{0l}/n_{0h}$. As a qualitative understanding, increasing the seed abundance shifts the model AGN LF upward, which must thus retreat leftward to meet data. This trend applies to both heavy- and light-seed BHs, whose bright-end LF slopes are mainly anchored by the quasar and JWST AGN data respectively.

Runs with different parameters yield converging heavy-seed LFs at $M_{1450}\lesssim-24$~mag and light-seed LFs at $-24~{\rm mag} \lesssim M_{1450}\lesssim-17~$mag. In the magnitude range $-20~{\rm mag} < M_{1450}<-16~{\rm mag}$, the model unobscured AGN LF is lower than the observed galaxy LF \cite{Bouwens2021} by a factor of 10--50, the difference shrinking toward the fainter end. Further, all model light-seed LFs still rise at $M_{1450}>-16~\rm mag$ until fainter than $-12~\rm mag$. Future deep AGN surveys will directly test our predictions. If the observed AGN LF starts to turn downward at $M_{1450}<-12~\rm mag$, then either substantially fewer light seeds than assumed actually become AGNs, or mergers between light BHs are more common, reducing the number density at the fainter end without overgrowing the most massive population.

While the AGN LF requires the light-seed BHs, the abundance of this population brings a large mass density: in all four runs, $\rho_{l,z5}>10^4~M_\odot$. This poses a potential tension with observation constraining the mass density budget of X-ray-selected AGNs. We will further discuss this issue in Section~\ref{subsec:mass_density}.

\subsection{Observed merger event rates}

\begin{figure*}
    \centering
    \includegraphics[width=0.75\textwidth]{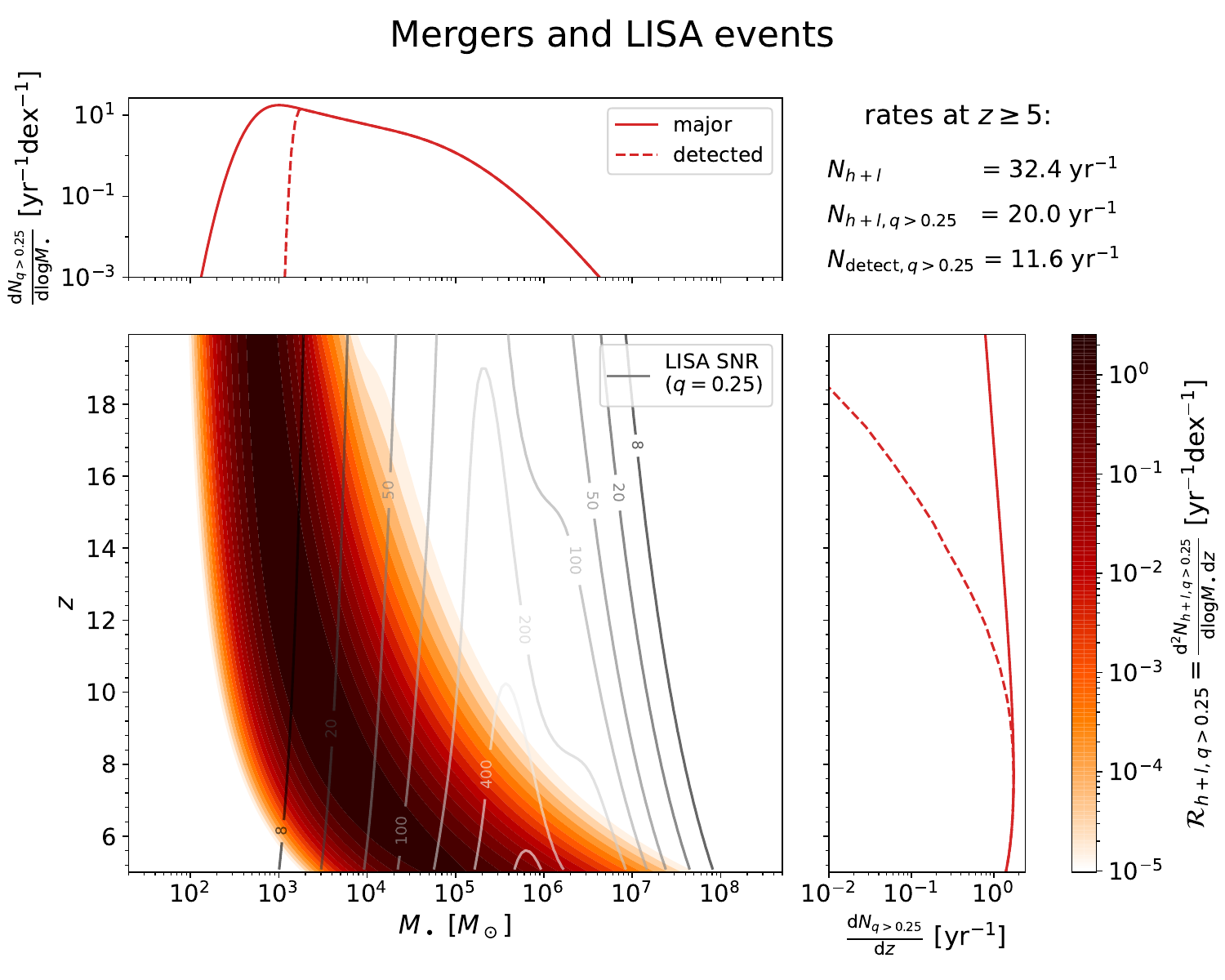}\vspace{-3mm}
    \caption{Differential merger rate corresponding to the heavy-and-light-seed model with $n_{0h}=10^{-3}~{\rm Mpc^{-3}}$ and $n_{0l}=10^0~{\rm Mpc^{-3}}$. \textit{Bottom left, main panel}: colored contours exhibit the major merger rate $\mathcal{R}_{h+l,q>0.25}$ as a function of the remnant BH mass $M_\bullet$ and redshift $z$. Greyscale contours track constant SNR from $q=0.25$ merger waveforms and the LISA noise curve. \textit{Top left}: the marginal major merger rate (solid) and the LISA-detectable rate (dashed) as a function of the remnant BH mass. The horizontal axis is shared with the main panel. \textit{Bottom right}: the marginal major merger rate (solid) and the LISA-detectable rate (dashed) as a function of redshift. The vertical axis is shared with the main panel.}
    \label{fig:rate2D}
\end{figure*}

\begin{figure*}
    \centering
    \includegraphics[width=0.75\textwidth]{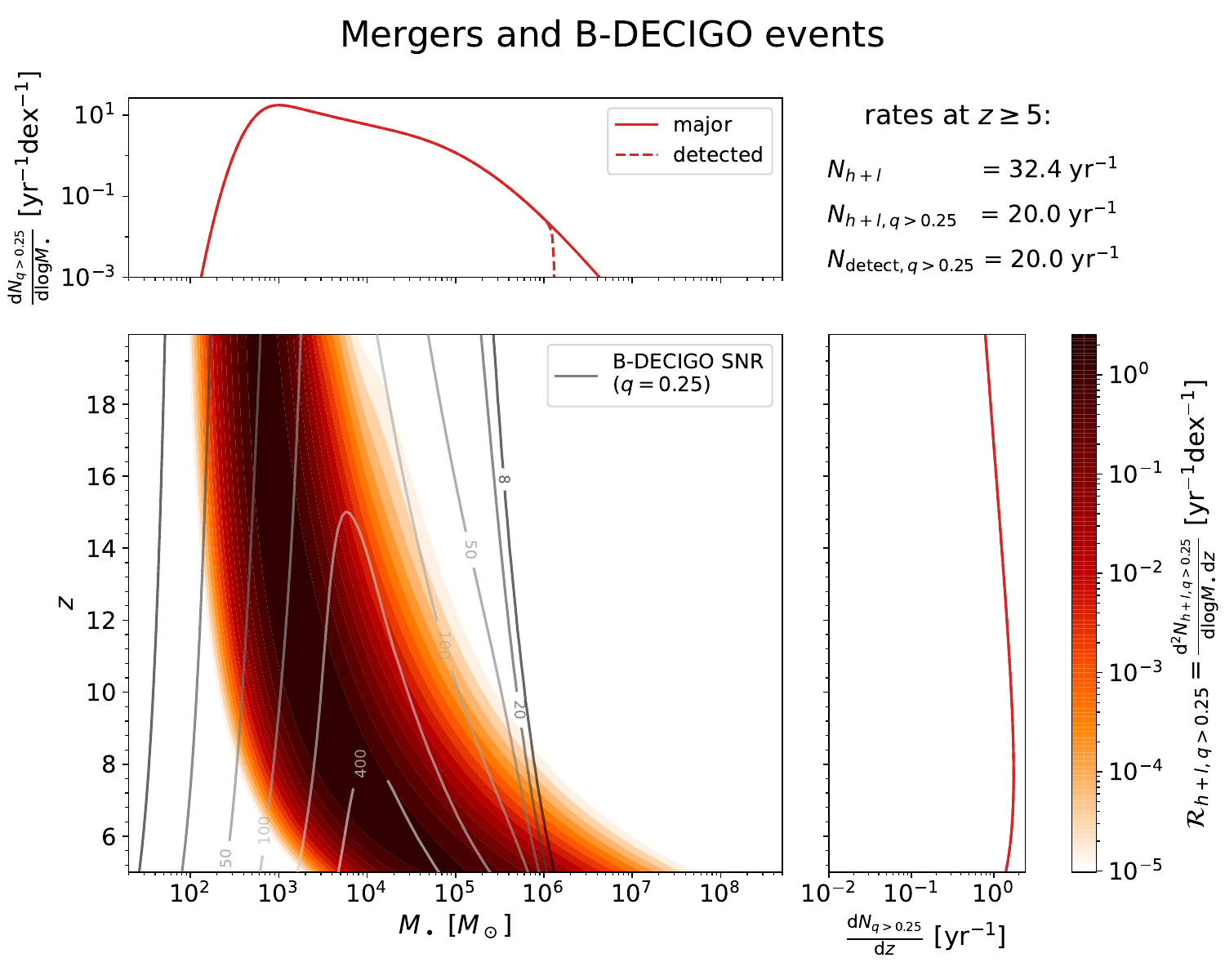}\vspace{-3mm}
    \caption{Same as Figure~\ref{fig:rate2D}, but for B-DECIGO. The solid and dashed curves overlap in the bottom-right panel.}
    \label{fig:rate2D_BDECIGO}
\end{figure*}

\begin{figure}
    \centering
    \includegraphics[width=0.47\textwidth]{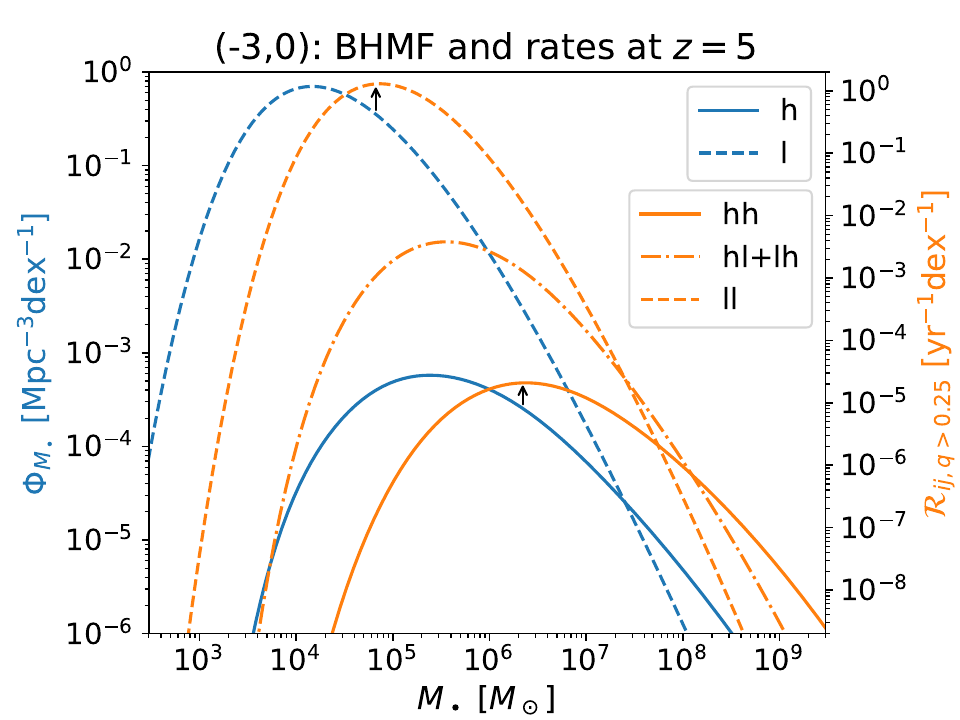}\vspace{-1mm}
    \caption{BHMF and merger rates in the run $n_{0h}=10^{-3}~{\rm Mpc^{-3}}$ and $n_{0l}=10^0~{\rm Mpc^{-3}}$ at $z=5$. \textit{Blue}: BHMF of the heavy- and light-seed BHs, the same as in Figure~\ref{fig:BHMF_hl}. \textit{Orange}: major merger rate per log mass per redshift. Black arrows indicate the mass $M$ satisfying $\Phi_{M_\bullet,i}(M/2)\propto M^{-1/2}$ where $i=h,l$. The peaks of heavy-heavy and light-light merger rates approximately match the black arrows and are heavier than the peaks of the BHMFs.}
    \label{fig:MF_vs_rate}
\end{figure}

Table~\ref{tab:merger_rate_hl} summarizes the number densities and merger rates of the two BH populations. The total and major merger rates for heavy-heavy, heavy-light, and light-light mergers satisfy $N_{hh}\ll N_{hl+lh}\ll N_{ll}$. This is expected from our sum-kernel formalism, which suggests $N_{ij}\propto \rho_i n_j$ at a given time in a given parameter set. The major merger rate accounts for a significant proportion in $N_{hh}$ and $N_{ll}$, but $N_{hl+lh,q>0.25}\ll N_{hl+lh}$ because heavy- and light-seed BHs are likely to have very different masses. The fact that the abundant light-seed BHs more readily undergo mergers also leads to a faster reduction in $n_l$ compared to $n_h$.  Since the best-fit merger timescales exceed the elapsed cosmic time, the heavy-seed BH number density (still calibrated with Equation~\ref{eq:number_decay}) undergoes negligible changes. 

The evolution of $n_h$ is independent of the light-seed population, which is not the case vice versa. Equation~(\ref{eq:number_decay_l}) implies that, if $n_h\ll n_l$ (valid in all four runs), increasing $\rho_h$ will slow down the decay in $n_l$. This explains the fact that the runs with $n_{0h}=10^{-2}~\rm Mpc^{-3}$ have shorter $\tau_m$ than $n_{0h}=10^{-3}~\rm Mpc^{-3}$, but $n_{l,z5}$ and $N_{ll}$ remain comparable for a fixed $n_{0l}$. Since all types of mergers obey the same coagulation kernel proportional to $\tau_m^{-1}\rho_h^{-1}$, a higher $\rho_h$ with a lower $\tau_m$ may give similar merger histories and number reduction of light-seed BHs.

Frequent merger rates enable statistical studies. In Figure~\ref{fig:rate2D}, we show the major merger rate distribution in the fiducial run $n_{0h}=10^{-3}~{\rm Mpc^{-3}}$ and $n_{0l}=10^{0}~{\rm Mpc^{-3}}$ as a function of the remnant BH mass and redshift down to $z=5$. Overall, the distribution tracks the MBH mass assembly process, the most frequent events occurring from $M_\bullet\sim10^3~M_\odot$ at $z=20$ through $M_\bullet\sim10^5~M_\odot$ at $z=5$. The maximum merger event rate density lies at $M_\bullet=2\times10^3~M_\odot$ at $z=11$. Similarly, in the top-left panel, the most frequent mergers correspond to remnant BH masses of $\sim10^3~M_\odot$. The major merger rate vs. mass distribution then follows a power law of $dN_{q>0.25}/d\log M_\bullet\propto M_\bullet^{-0.5}$ until breaking at the point $(4\times10^4~M_\odot,3~\rm yr^{-1})$. In the bottom-right panel, the redshift evolution of the merger rate forms a moderate peak at $z=8$ due to the combined effect of the nonlinear redshift-time correspondence ($dt/dz$ decreases with $z$) and the MBH number density reduction. 

Overplotted in the main panel in Figure~\ref{fig:rate2D} are the constant SNR contours observed with LISA, evaluated from $q=0.25$ merger waveforms and the LISA noise curves. The four-year galactic foreground noise \cite{Babak2021} is considered in the calculation. For simplicity, only major mergers with $\rm SNR\geq SNR_{thresh} = 8$ are assumed to be detectable. In the visualized case, LISA can identify more than half of the major mergers at $5<z<20$, but the SNR threshold lies close to the location of the maximum merger rate density discussed above. Mergers of $M_\bullet\lesssim10^3~M_\odot$ at $z\gtrsim 6$ will be missed, which requires observatories sensitive to higher frequencies. 

In Figure \ref{fig:rate2D_BDECIGO}, we repeat the calculation with the noise curve of B-DECIGO. This observatory will focus on the dHz band and thus cover a lighter mass range than LISA. Almost all the predicted major mergers fall within $\rm SNR\geq8$, allowing for a full characterization of early MBH mergers.

\subsection{Comparing BHMF with merger rates}

Given sufficient merger events, one may hope to extract demographic information from the merger rates, e.g., finding the most abundant BH mass. Our formulation predicts that the peak of the BHMF is likely lighter than the mass producing the most frequent mergers since the coagulation kernel is positively correlated with mass. 

Figure~\ref{fig:MF_vs_rate} illustrates this with the run $n_{0h}=10^{-3}~{\rm Mpc^{-3}}$ and $n_{0l}=10^0~{\rm Mpc^{-3}}$. At $z=5$, the light-seed BHMF peaks at $1.5\times10^4~M_\odot$, whereas the light-light major merger rate peaks at $7.4\times10^4~M_\odot$. Similarly, the most abundant heavy-seed BHs weigh $2.4\times10^5~M_\odot$, compared to $2.4\times10^6~M_\odot$ for the heavy-heavy major merger rate. 

One may understand this difference analytically. The merger rate per unit logarithmic mass and unit redshift, $\mathcal{R}_{ij}$, is proportional to $\psi_{ij}$ (Equation~(\ref{eq:observed_rate})). In Equation~(\ref{eq:psi_ij_def}), we now insert a delta-function $\delta(m-M/2)$ into the integral to account for equal-mass mergers only, and thus $\psi_{ii}(M)\propto M\Psi^2_{M_\bullet,i}(M/2)$ (the sum kernel contributes to a factor of $M$). Therefore, the mass $M$ giving $\Psi_{M_\bullet,i}(M/2)\propto M^{-1/2}$ approximately matches the maximum of $\psi_{ii}$. The black arrows in Figure~\ref{fig:MF_vs_rate} point from such locations vertically, and indeed, the arrow heads are close to the peaks of $\psi_{ii}$.

In the above example, the most abundant mass of the BHMF and the merger rate differ by a factor of 5 (10) for light- (heavy-) seed BHs. In general, as long as the coagulation kernel is positively correlated with the BH mass, the difference is at least a factor of 2. Because mergers preferentially occur between high-mass BHs, the GW event demographics are biased toward the massive end and require careful statistical consideration to infer the underlying BHMF. 

\section{Discussion}
\label{sec:discussion}

\subsection{Delayed mergers}
\label{subsec:delay}
\begin{figure*}
    \centering
    \includegraphics[width=0.99\textwidth]{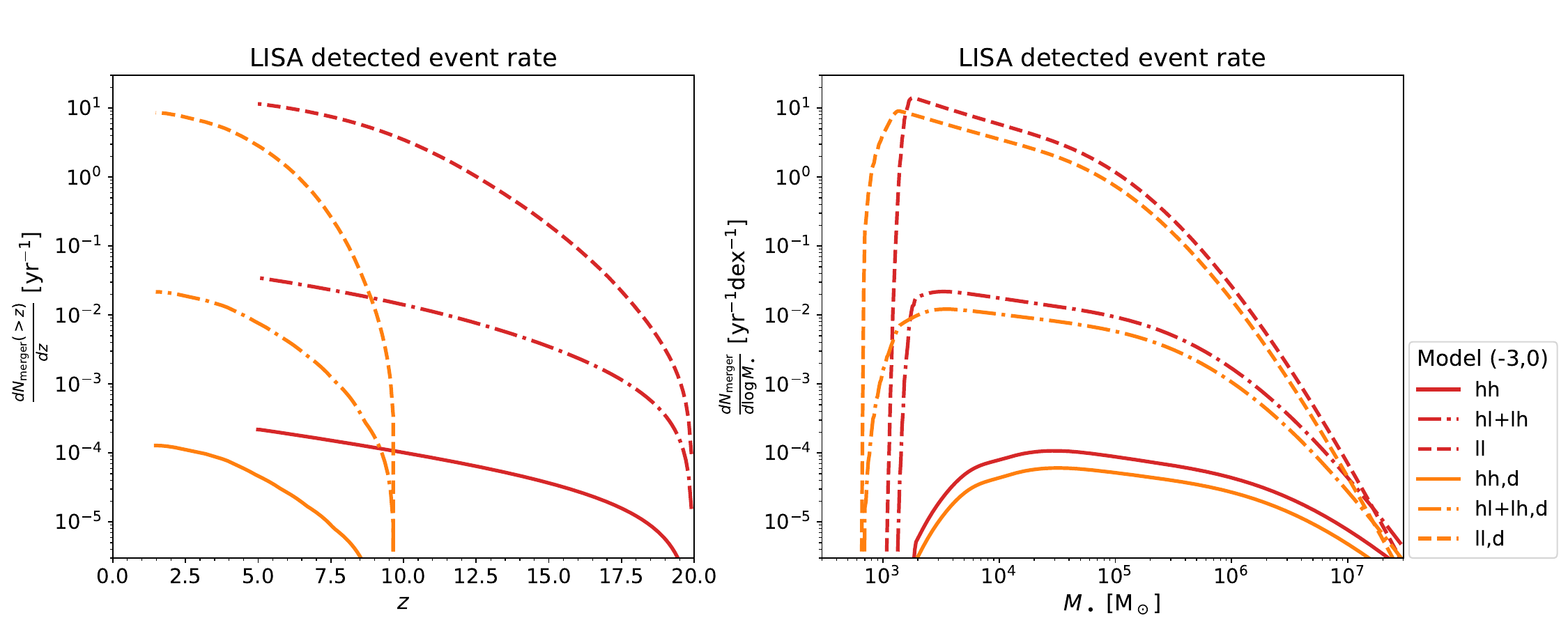}
    \caption{Differential major merger rate detected by LISA in the run with $n_{0h}=10^{-3}~{\rm Mpc^{-3}}$ and $n_{0l}=10^0~{\rm Mpc^{-3}}$, decomposed into heavy-heavy (hh), heavy-light (hl+lh), and light-light (ll) mergers. Red curves are measured from the fiducial model. Orange curves (with the legend ``d") represent post-processed delays with a delayed timescale $10^{2.5}~{\rm Myr}<t_{\rm delay}<10^{3.5}~{\rm Myr}$. \textit{Left}: cumulative rate as a function of redshift, similar to the bottom-right panel in Figure~\ref{fig:rate2D}. \textit{Right}: rate as a function of the remnant BH mass, similar to the top-left panel in Figure~\ref{fig:rate2D}. }
    \label{fig:delay}
\end{figure*}

Thus far, our analysis assumes a constant $\tau_m$ to characterize the number density reduction of MBHs. However, constraints from AGN LFs do not rule out the possibility that two BHs orbiting each other could mimic a single BH powering an AGN (due to the lack of spatial resolution to separate a dual AGN and periodic signatures indicative of a binary BH system). Their orbital decay may be delayed before reaching the final merger stage, which our model using an average constant $\tau_m$ does not account for. The delay of BH mergers is influenced by various environmental factors within the galactic nucleus, introducing significant uncertainty in theoretical models of GW event rates \cite{AmaroSeoane2023}. Optimistic estimates based on BH-disk interactions suggest a delay timescale as short as $\sim10~\rm Myr$ \cite{Haiman2009}, while more pessimistic scenarios propose delays exceeding a Hubble time. In this subsection, we briefly discuss how these potential delays might affect the predicted merger rates.

To account for delay effects occurring at unresolved nuclear scales -- resulting from inefficient energy and angular momentum transport from the binary BH to the surrounding stars and gas -- we apply a time shift to the same merger rate distribution as obtained in the fiducial run. We adopt a log-uniform distribution of the delay time $10^{2.5}~{\rm Myr}<t_{\rm delay}<10^{3.5}~{\rm Myr}$, comparable to the light-seed scenario with delays but without supernova feedback as discussed in Ref.~\cite{Barausse2020}. We assume that BHs pause accretion when subject to this delay.

Figure~\ref{fig:delay} presents the major mergers detectable by LISA in the run with $n_{0h}=10^{-3}~{\rm Mpc^{-3}}$ and $n_{0l}= 10^0~\rm Mpc^{-3}$ both with (orange) and without delays (red). The left panel shows the cumulative merger rate by a redshift of $z$ and the right panel shows the total detected rate differentiated by the logarithmic remnant BH mass. In each panel, we distinguish between the three types of mergers: light-light (dashed), light-heavy (dot-dashed), and heavy-heavy seed BHs (solid). We first describe the results without delays. The cumulative detection rate is dominated by mergers between light-seed BHs (see also Figure~\ref{fig:MF_vs_rate}). This rate exceeds $1~{\rm yr}^{-1}$ at redshifts $z\leq12.5$ and saturates at $11.6~{\rm yr}^{-1}$, whereas the event rates involving heavy-seed BHs only amount to $<10^{-1}~\rm yr^{-1}$. In the right panel, BH mergers with remnant masses of $M_\bullet \sim 2\times 10^3~\msun$ are most likely to be detected. Moreover, detections at higher masses will constrain the shape of the merger-rate mass function, which is predicted to follow a double power-law form with indices $M_{\bullet}^{-0.5}$ and $M_{\bullet}^{-2.5}$ and a characteristic mass scale of $M_\bullet \simeq 10^5~\msun$. Note that within the four-year LISA operation period, the shape of the merger-rate mass function is expected to be well constrained for $10^3~\msun<M_\bullet < {\rm a~few}\times 10^5~\msun$. In contrast, the rate--mass distributions of mergers involving heavy-seed BHs exhibit shallower slopes, though these are unlikely to be detectable. 

We now turn to the model with delays\footnote{The curves with delays only consider mergers that would occur at $z\geq5$ if without delays.}. The event rates are still dominated by light-light mergers. The minimum delay time of $10^{2.5}~\rm Myr$ prevents mergers from occurring at $z>10$. At lower redshifts, the detected event rate accumulates and reaches $1~\rm yr^{-1}$ at $z=6.4$, or 500 Myr later than the no-delay case. The saturated detection rate of $8.6~\rm yr^{-1}$ is slightly lower than our no-delay case as the comoving volume per unit cosmic time decreases at lower redshifts, though this is partially compensated by LISA becoming more sensitive at lower redshifts and capable of detecting a larger proportion of the mergers. Since our assumed maximum delay time of $10^{3.5}~\rm Myr$ is significantly shorter than the time interval between $z=5$ and the present universe, those high-redshift mergers are only postponed but not stalled, i.e., the BH number density reduction at the terminal redshift is unchanged. A maximum delay time comparable to or longer than a Hubble time would instead cause some BHs never to merge and significantly reduce the detected GW event rate.

On the other hand, the mass distributions of event rates with and without delays are similar in the double-power-law profile at remnant masses $M_\bullet>10^3~M_\odot$. The model with delays predicts a lower contribution on the heavy side, mainly owing to the reduced differential comoving volume at low redshifts, and a higher contribution for $M_\bullet\lesssim10^3~\msun$ due to these mergers occurring at closer luminosity distances and thus more readily detectable by LISA.

\subsection{Comparison with other works}

\begin{figure*}[ht]
    \centering
    \includegraphics[width=0.99\textwidth]{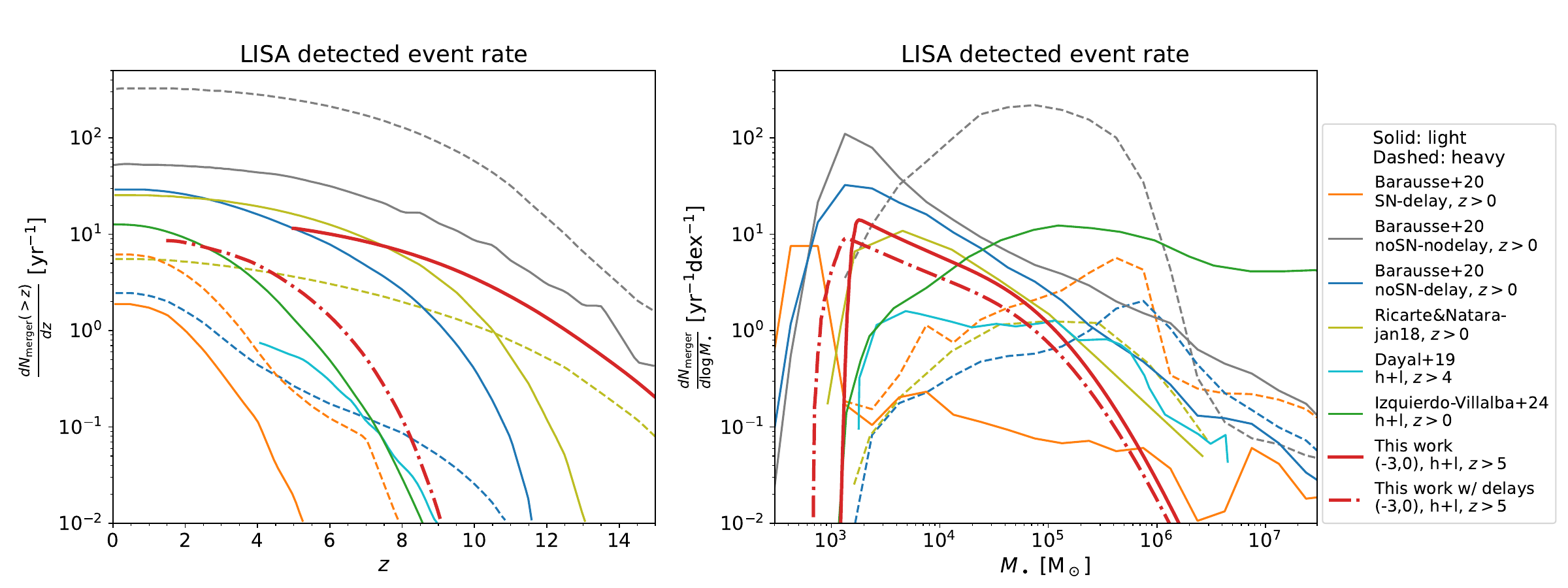}
    \caption{Differential major merger rate detected by LISA in different works, similar to Figure~\ref{fig:delay}. Red solid curves present the total (h+l) rate in this work with $n_{0h}=10^{-3}~{\rm Mpc^{-3}}$ and $n_{0l}=10^0~{\rm Mpc^{-3}}$. Red dot-dashed curves represent the total (h+l) rate with post-processed delays. \textit{Left}: cumulative rate as a function of redshift. Adapted from Ref.~\cite{AmaroSeoane2023}. This work's results resemble previous literature with or without delays. \textit{Right}: rate as a function of the remnant BH mass.}
    \label{fig:rate_literature}
    \vspace{2mm}
\end{figure*}

Many previous studies have modeled detailed physical scenarios, such as BH seeding and BH sinking processes through gravitational interactions with surrounding gas and stars, and their influence on the merger rate, based on dark-matter halo merger trees (e.g., \cite{Barausse2012,Ricarte2018,Bonetti2019,Dayal2019,Barausse2020,Izquierdo-Villalba2024}). However, our approach differs: the merger timescale, $\tau_m$, is a free parameter constrained by the unobscured AGN LF data. Mergers modify the MBH demographics, characterized by the number density at the zeroth order and the BHMF shape in more detail. The demographics, in turn, provide clues about the merger rate. We directly calculate BH properties without assuming a specific relationship between the BHs and their host galaxies or dark-matter halos, and our method is insensitive to uncertainties of delays at resolved spatial scales.

In Figure~\ref{fig:rate_literature}, we compare our detection rate distributions of all merger types with those from previous studies \cite{Ricarte2018,Barausse2020,Dayal2019,Izquierdo-Villalba2024}\footnote{Ref.~\cite{Dayal2019} reported the merger rate mass function depending on the redshifted mass and used an SNR threshold of 7 instead of 8. When plotting the right panel, we shift their curve left by 7 to approximate the intrinsic BH mass since their detected merger rate peaked at $z\sim6$. Ref.~\cite{Ricarte2018} reported the same function depending on the chirp mass, which we multiply by 3 to approximate the remnant mass. The left panel shows the accurate results coordinated in Ref.~\cite{AmaroSeoane2023}. Ref.~\cite{Izquierdo-Villalba2024} reported the total merger rate as a function of redshift and the detected rate as a function of mass with an SNR threshold of 10. We present their results without modification.}. The first two works cited above considered two distinct BH seeding models: light-seed (solid) and heavy-seed BHs (dashed), in the sense of Pop~\RNum{3} remnants versus direct-collapse BHs. The third work combined the two seeding channels (to which this study is similar); the last work used a mixed formalism of heavy- and light-seed BHs and single-mass seed BHs. We use solid curves in the figure to represent their merger rates. Across the literature, even in a single study, theoretical event rate predictions can vary by orders of magnitude depending on the physical assumptions and techniques used (see also discussion in Ref.~\cite{AmaroSeoane2023}). 

The redshift trend of the detection rate in our fiducial case closely resembles previous models that do not account for delays, including Ref.~\cite{Ricarte2018} and the no-delay, no-supernova (SN) scenario in Ref.~\cite{Barausse2020}. In contrast, the models with delays primarily at larger scales but not in the nucleus (Refs.~\cite{Dayal2019} and the scenarios in Ref.~\cite{Barausse2020} with delays) typically show detection rates that increase significantly later, after the initial seeding of BHs, and their cumulative rates at a given redshift are substantially lower than their no-delay counterparts (e.g., the blue vs. gray curve). The model in Ref.~\cite{Izquierdo-Villalba2024} considers delays on multiple spatial scales and exhibits a similar redshift distribution. Our model with delays at unresolved nuclear scales agrees with the features above. The mass function of merger rates in our no-delay model is comparable to the light-seed, no-SN cases in Ref.~\cite{Barausse2020} and the light-seed case in Ref.~\cite{Ricarte2018}. Our curves drop quickly at both the light and heavy ends of the mass spectrum, mainly because our analysis do not cover $z<5$, where LISA would have access to a broader detectable mass range. We have discussed in Section~\ref{subsec:delay} that the delays in our model do not strongly modify the profile of the merger-rate mass function, which agrees with, e.g., the comparison between the noSN-nodelay and noSN-delay cases in Ref.~\cite{Barausse2020}. The mass dependence of the event rate in Ref.~\cite{Izquierdo-Villalba2024} shows a shallow slope at the massive end, compared to other studies, to account for the large amplitude of the GW background measured with PTAs.

Comparison requires caution. We consider co-existing and co-evolving heavy- and light-seed BHs, which formed in early cosmic regions biased at different extents. Our heavy-heavy merger rates fall below the boundary of Figure~\ref{fig:rate_literature} (but see Figure~\ref{fig:delay}). Our approach differs from, for instance, Ref.~\cite{Barausse2020}, where only one type of seed was considered in each model setup and $\sim 5-6$ orders of magnitude higher event rates were predicted.  On the other hand, Ref.~\cite{Dayal2019} studied multiple populations and found light-light mergers dominating the event rates, similar to our findings. Compared to their work, ours originally shows that multiple MBH seeding channels are not just theoretically expected \cite{Inayoshi2020} but also observationally preferred by JWST AGN observations.

\subsection{GW Localization and observation of merger afterglows}

A substantial GW-event detection rate potentially enables multi-messenger observations of these MBH mergers. Combining EM and GW signals is expected to reveal rich insights into the physics and environment of the evolving BH \cite{AmaroSeoane2023}. In the following, we estimate the opportunity of unambiguously identifying a merger's possible afterglow at $z>5$, i.e., observing EM signals after the GW event. This will utilize the full inspiral, merger, and ringdown signals in GWs to minimize localization errors.

If a BH merger leads to a GW detection with uncertainties of the remnant mass $\Delta\log M_\bullet$, sky localization $\Delta\Omega$ (in units of deg$^2$), and luminosity distance $\Delta\log D_L$, the number of MBHs within this parameter error box is given by 
\begin{equation}
    \mathcal{N}_\bullet = \Phi_{M_\bullet}\frac{d^2V_C}{d\log D_L d\Omega}\Delta\Omega \Delta \log D_L \Delta\log M_\bullet\,,
\end{equation}
where $d^2V_C/d\log D_L d\Omega=1.0\times10^8~\rm Mpc^3\,deg^{-2}$ at $z=5$. Ref.~\cite{Mangiagli2020} calculated the LISA parameter estimation errors of BH mergers up to $z=4$, which we extrapolate to $z=5$ and find the optimal error box at $M_{\bullet}=6\times10^5~M_\odot$ for major mergers. The error of each parameter is calculated as $\Delta\Omega=1.1~\rm deg^2$, $\Delta\log D_L=4.0\times10^{-3}$, and $\Delta\log M_\bullet=1.3\times10^{-3}$, where the full GW waveforms including inspiral, merger, and ringdown phases (PhenomC \cite{Santamaria2010}) are taken into account\footnote{Ref.~\cite{Mangiagli2020} calculated the uncertainties in the inspiral stage. They then scaled the sky localization uncertainty by $(\rm SNR)^{-2}$ and the luminosity distance uncertainty by $(\rm SNR)^{-1}$ to account for the full waveforms, where SNR is the accumulated GW signal-to-noise ratio; see their Equations~(13) and (14). We similarly scale their relative mass uncertainty in the inspiral stage by $(\rm SNR)^{-1}$ (accurate for post-Newtonian inspiral signals \cite{Cutler1994}, but only approximate for the full waveforms) to obtain the full-waveform uncertainty.}.
Since the total MBH abundance at $M_\bullet=6\times10^5~M_\odot$ is $\Phi_{M_\bullet,l}=2.5\times10^{-2}~\rm Mpc^{-3}\,dex^{-1}$ (see Figure~\ref{fig:BHMF_hl}), we thus obtain $\mathcal{N}_\bullet=14~[\Delta \Omega\Delta\log D_L/(4.4\times 10^{-3}~{\rm deg}^2)]$. 

A more observationally relevant estimate is the number of AGNs within sky localization and redshift errors, which will become candidates to associate with the merger event. This is given by
\begin{equation}
    \mathcal{N}_{\rm AGN} = \frac{d^2V_C}{d\log D_Ld\Omega} \Delta\Omega \Delta\log D_L \int_{M_{1450}^{\rm min}}^{M_{1450}^{\rm max}}\Phi_{1450}d M_{1450}\,,
\end{equation}
where $M_{1450}^{\rm min}$ and $M_{1450}^{\rm max}$ give the physically plausible magnitude range of the merger afterglow. If we assume that the merger remnant BH launches super-Eddington accretion, then $M_{1450}^{\rm min}=-\infty$ and $M_{1450}^{\rm max}=-18.2$ mag, given by $L_{\rm Edd}=7.5\times10^{43}~\rm erg\,s^{-1}$ at the mass $M_\bullet=6\times10^5~M_\odot$. This implies $\mathcal{N}_{\rm AGN}=73~[\Delta \Omega\Delta\log D_L/(4.4\times 10^{-3}~{\rm deg}^2)]$, higher than $\mathcal{N}_\bullet$ because the Eddington ratio of a post-merger BH is highly uncertain.

The result of $\mathcal{N}_{\rm AGN}\gg1$  challenges matching a LISA-detected event at $z\geq5$ to a unique EM target. Note that this estimation is based on the BHMF and the unobscured AGN LF where our models with different $n_{0h}$ and $n_{0l}$ converge within an order of magnitude. However, more accurate parameter estimation is possible. Regarding the source, a GW event with a low inclination angle and high ecliptic latitude may achieve an order of magnitude smaller $\Delta\Omega$ than the above-reported average result \cite{Mangiagli2020}. Regarding the detector, joint detection with a network of GW observatories may improve the sky localization by a factor of a few to several orders of magnitude (e.g., LISA with B-DECIGO, \cite{Grimm2020}; with TianQin, \cite{TorresOrjuela2024}; with Taiji, \cite{Wang2020}), depending on the detector design. Enhanced precision in sky location and distance measurements, $\Delta\Omega\Delta\log D_L\leq6.0\times10^{-5}~\rm deg^2$, will filter out AGNs irrelevant to the merger and enable deep subsequent follow-up observations.

\subsection{Evolution of MBH mass density}

\label{subsec:mass_density}
\begin{figure}
    \centering
    \includegraphics[width=0.50\textwidth]{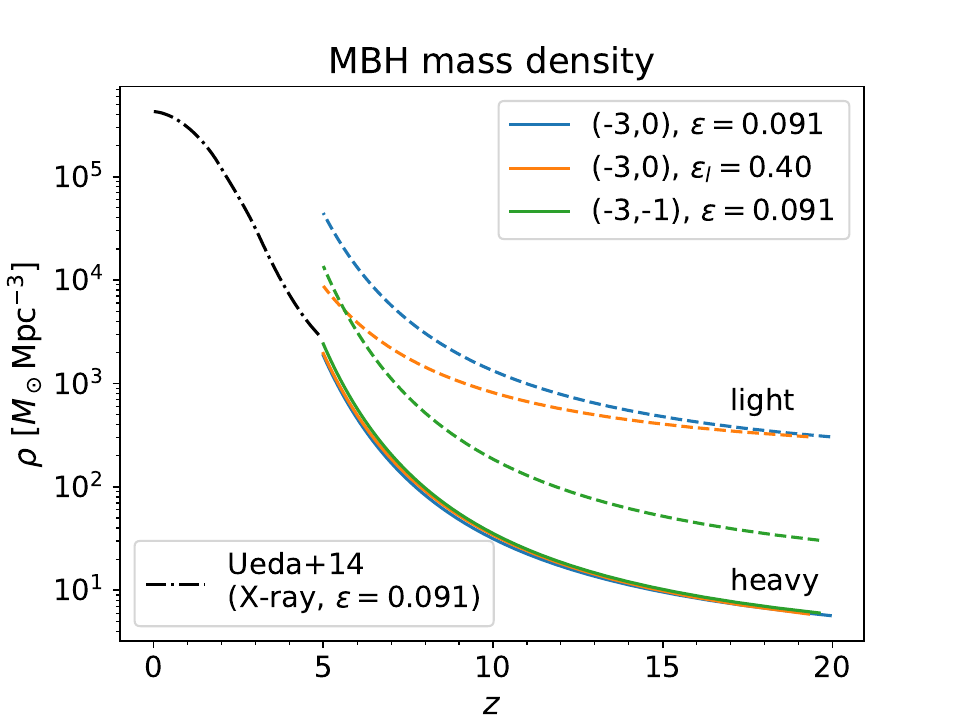}
    \caption{MBH mass density evolution from the fiducial model with $n_{0h}=10^{-3}~{\rm Mpc^{-3}}$ and $n_{0l}=10^0~{\rm Mpc^{-3}}$ (blue), an alternative model with increased radiative efficiency of light-seed BHs, $\epsilon_l=0.40$ (orange), and an alternative model with $n_{0l}=10^{-1}~\rm Mpc^{-3}$ (green). Solid (dashed) curves correspond to the heavy- (light-) seed BHs. The evolution inferred from X-ray AGNs \cite{Ueda2014} assuming a radiative efficiency $\epsilon=0.091$ is presented as the dot-dashed curve. The fiducial model overshoots observation at $z\sim5$ by an order of magnitude, which may be alleviated by increasing $\epsilon_l$ or decreasing $n_{0l}$. }
    \label{fig:mass_density}
\end{figure}

The evolution of the MBH total mass density, driven solely by accretion, is directly linked to the luminosity of AGNs. With an assumed radiative efficiency $\epsilon\sim0.1$, the total AGN luminosity integrated over cosmic time agrees with the dynamically measured mass density of local MBHs. This is known as the Soltan argument \cite{Soltan1982}. Following this idea, X-ray observations have constrained the mass budget of MBHs to high redshifts ($z\lesssim5$, \cite{Hopkins2007,Ueda2014}). 

Figure~\ref{fig:mass_density} presents the MBH mass density evolution calculated using the best-fit parameters in our model with $n_{0h}=10^{-3}~{\rm Mpc^{-3}}$ and $n_{0l}=10^0~{\rm Mpc^{-3}}$ (blue curves). This is compared to the $z\lesssim5$ result inferred from the X-ray AGN luminosity assuming $\epsilon=0.091$ (\cite{Ueda2014}, the plot here adjusted for the different values of the assumed radiative efficiency). The heavy-seed population is consistent with constraints from the X-ray, in agreement with previous works \cite{Li2024}. However, the light-seed population at $z\sim5$ has a mass density exceeding observation by an order of magnitude.

One potential alleviation is that the light-seed BHs have radiative efficiency $\epsilon_l>0.1$, which requires less $\rho_{l,z5}$ to account for the JWST-selected AGN abundance. \cite{Inayoshi2024} argued for a high radiative efficiency for the dust-reddened, broad-line AGNs discovered by JWST based on their accreted mass density at $5<z<8$. To explore a similar scenario for unobscured AGNs, we perform an additional MCMC fitting with $n_{0h}=10^{-3}~{\rm Mpc^{-3}}$ and $n_{0l}=10^0~{\rm Mpc^{-3}}$ but with the light-seed BH radiative efficiency $\epsilon_l$ modified to 0.4, close to the theoretical limit for maximally rotating BHs. The best-fit result is also presented in Figure~\ref{fig:mass_density} (orange). Compared to the fiducial model, this setup matches the observed LF slightly better ($\ln P_{\rm post}=-43$), with $\rho_{l,z5}$ reduced by a factor of 5. However, the total mass density remains in tension with the X-ray observations. 

Alternatively, we may speculate that fewer light seeds participate in the MBH evolution than we have estimated. We thus run another MCMC with $n_{0h}=10^{-3}~{\rm Mpc^{-3}}$ and $n_{0l}=10^{-1}~{\rm Mpc^{-3}}$ and with fiducial radiative efficiencies, with the best fit shown in Figure~\ref{fig:mass_density} (green). The maximum posterior probability ($\ln P_{\rm post}=-66$) is mildly worse than fiducial, but $\rho_{l,z5}$ reduces by a factor of 3. The reduction factor in the above scenario is less than the naive expectation of 10 (from the change of the initial number density) as MCMC tunes the parameters to fit the AGN LF.

The discrepancy between our model's total BH mass density and the integrated mass density inferred from X-ray-selected AGNs at $z\sim5$ is a consequence of the abundance of the JWST-selected unobscured AGN candidates, which are one to two orders of magnitude more numerous than the X-ray bright AGN populations at $z\gtrsim 4-5$ of comparable bolometric luminosities (noted in, e.g., \cite{Harikane2023}; see also \cite{Inayoshi2024}). A possible explanation for the difference in abundance is that previous X-ray surveys at $4<z<5$ may have missed AGNs that are X-ray faint, a common feature in unobscured AGNs identified with JWST (e.g., \cite{Yue2024,Maiolino2024}). However, to comply with Soltan's argument, the growth of this X-ray faint population's mass density at lower redshifts must slow down and become surpassed by that of the X-ray bright BHs, unless both populations have radiative efficiencies higher than $\sim0.1$. The decelerated mass density growth could be due to weakened accretion or reduced membership, the latter by mergers with X-ray bright populations or by unknown processes enhancing X-ray emission. 
Future deep surveys on low-luminosity AGNs in the redshift range $0<z<5$ will be essential in establishing representative MBH samples, connecting the present-day BHs and their progenitors in the first billion years of the universe.

\section{Summary}
\label{sec:summary}

The evolutionary history of MBHs is being unraveled through EM observations and will be further elucidated by GW probes. In this study, we model the BH mass assembly process in light of the new JWST-identified AGNs to investigate BH formation channels and predict GW events. We consider two types of seed BHs: heavy seeds ($M_\bullet \sim 10^{2-5}~\msun$) 
formed in rare and overdense cosmic regions, and light seeds ($M_\bullet \sim 10^{1-3}~\msun$) formed as stellar remnants in commoner dark-matter halos.
The BHs grow through episodic accretion and mergers, which are constrained by the LF data of quasars and JWST-selected AGN candidates at $z\sim 5$.
This work builds upon the model presented in Ref.~\cite{Li2023}, incorporating new elements including multiple seeding channels, mergers, 
and JWST-selected AGN candidate data. The model aims to contextualize future multi-messenger observations of MBHs within their mass assembly history for $5 \lesssim z \lesssim 20$.
We highlight our results as follows:

\begin{enumerate}
    \item While reproducing the $z\sim5$ quasar LF, models with heavy-seed BHs alone have difficulty explaining the JWST-selected faint AGN candidates (Figure~\ref{fig:QLF_heavy}). Our best fits with the seed abundance $n_{0h}=10^{-3}~{\rm Mpc}^{-3}$ ($10^{-2}~{\rm Mpc}^{-3}$) fall below the data by a factor of $\gtrsim10$ ($\gtrsim5$).
    \item The combined heavy- and light-seed BH population explains the observed AGN LF throughout available magnitudes (Figure~\ref{fig:QLF_hl}). Quasars with $M_{1450}\leq-22~\rm mag$ are mainly attributed to heavy-seed BHs, but most fainter, JWST-selected AGN candidates may originate from light seeds.
    \item Both the total merger rate and the merger rate involving only heavy-seed BHs are subject to model-independent upper bounds (Equations~\ref{eq:rate_upper_limit} and \ref{eq:rate_upper_limit_hl}) and further restricted by the AGN LF. We predict a heavy-heavy major merger rate of $\lesssim10^{-1}~\rm yr^{-1}$. On the other hand, light-seed BHs give merger rates $>10^1~\rm yr^{-1}$ mainly due to their abundance, with major mergers comprising a significant fraction (Table~\ref{tab:merger_rate_hl}) and being readily detectable by space-based GW observatories such as LISA and B-DECIGO (Figures~\ref{fig:rate2D} and \ref{fig:rate2D_BDECIGO}). The remnant BH mass distribution of the detected GW events is predicted to follow a double power-law form with $M_{\bullet}^{-0.5}$ and $M_{\bullet}^{-2.5}$ and a characteristic mass scale of $M_\bullet \simeq 10^5~\msun$. Within the LISA operation period, the mass function shape is expected to be well constrained for $10^3~\msun<M_\bullet <{\rm a~few}\times 10^5~\msun$ (Figure~\ref{fig:delay}). 
    \item If a merger occurs with a remnant BH mass $M_\bullet=6\times10^5~M_\odot$, where LISA achieves optimal SNR, the number of AGNs as afterglow candidates within the parameter estimation error box is $\mathcal{N}_{\rm AGN}=73~[\Delta \Omega\Delta\log D_L/(4.4\times 10^{-3}~{\rm deg}^2)]$. Precise sky localization and distance measurement of the GW events, possibly facilitated by preferred source direction and multi-detector GW observation, may enable electromagnetic identification of mergers at $z\geq5$ and subsequent multi-messenger follow-up observations.
\end{enumerate}

Interpreting our results comes with several caveats. Firstly, our merger prescription is simplistic (Section~\ref{subsec:merger_coagulation}). Without a more complicated parameter design, our model predicts trends similar to previous works without detailed physics such as delays and feedback. Secondly, all our best-fit merger timescales are longer than $10^4~\rm Myr$. This tends to be pessimistic because $\tau_m$ is the only parameter for mergers, and MCMC may prefer the more flexible accretion channel tuned by four parameters. Still, our analytical upper bounds for the merger rates apply to all merger timescales. Thirdly, the JWST AGN candidate data adopted here await confirmation and provide upper bounds of the real unobscured AGN LF, although the data are consistent with those based on broad-line AGN samples compiled in Refs.~\cite{Harikane2023,Maiolino2023}. The obscuration fraction of these JWST-selected AGNs is also uncertain as the value adopted in this work is inferred from X-ray-bright AGNs at $z\leq5$. The intrinsic AGN LF at the faint end will be refined by future spectral follow-up observations and multi-wavelength surveys.

\begin{acknowledgments}
We thank the anonymous referee for constructive comments that helped to improve this work.
We greatly thank Wenxiu Li for sharing their code to model BH accretion and compare theory to data.
We also wish to thank Xian Chen, Jingsong Guo, and Priyamvada Natarajan for constructive discussions and Pau Amaro-Seoane for help with figure adaptation. 
K.~I. acknowledges support from the National Natural Science Foundation of China (12073003, 12003003, 11721303, 11991052, 11950410493), 
and the China Manned Space Project (CMS-CSST-2021-A04 and CMS-CSST-2021-A06). 
MCMC computation in this work is supported by the High-Performance Computing Platform of Peking University. 
\end{acknowledgments}

\appendix

\section{Physical interpretation and simplistic models of the coagulation kernel}
\label{app:kernel_models}
Here, we describe two models to interpret the coagulation kernel qualitatively and justify its positive mass dependence. Before proceeding, we note that the coagulation kernel $K(M,m,t)$ has a dimension of $\rm Length^3\times Time^{-1}$. This implies the physical interpretation of the kernel as $K=\sigma v_{\rm rel}$, i.e., the product of the BH collisional cross-section and the relative velocity. 

Firstly, a simplistic treatment of two merging BHs assumes two-body Newtonian interactions, and a merger likely occurs when the BH distance is sufficiently close such that general relativistic corrections become significant. Each BH is thus crudely analogous to a hard sphere with radius $R=\beta R_{\rm Sch} = 2\beta GM/c^2$, where $\beta$ is a dimensionless factor and $R_{\rm Sch}$ is the Schwarzschild radius. For BHs with equal masses, as they approach each other with an impact parameter $b$ and relative velocity at infinity $v_{\rm rel}$, their closest distance $r_p$ satisfies
\begin{equation}
    \pi b^2 = \pi r_p^2\left(1+\frac{4GM}{v_{\rm rel}^2r_p}\right)\,.
\end{equation}
The enlarged $b$ relative to $r_p$ is known as gravitational focusing. Substituting $r_p = 2\beta GM/c^2$ into the above formula gives the collisional cross-section. If $\beta$ is of order unity and $v_{\rm rel}\ll c$, then
\begin{equation}
    \sigma = \pi b^2 \simeq \frac{8\pi \beta G^2}{c^2}\frac{M^2}{v_{\rm rel}^2}\,. \label{eq:cross_section_hardsphere}
\end{equation}

The above model idealizes BH mergers regardless of their host galaxies. Alternatively, we model that BH mergers are related to the collisional cross-section of their host galaxies, assuming the BH mergers closely follow the galaxy mergers. Specifically, we treat galaxy mergers using a tidal impulse approximation. Between the two progenitor galaxies, one is considered as a gravitational perturber to the other. Assuming that the impact parameter is much larger than the characteristic galaxy size (the tidal approximation) and that the encounter timescale is much shorter than the galaxy crossing timescale (the impulse approximation), one expands the perturbed gravitational potential to second order to derive the decrement of the intergalactic potential energy \cite{Spitzer1958}
\begin{equation}
    \Delta E = \frac{4G^2M_p^2M_gR_g^2}{3R_p^4v_p^2}\,,
\end{equation}
where $M_p$ and $M_g$ are the masses of the perturber and the perturbed galaxy, $R_g$ is the root-mean-squared radius of the perturbed galaxy, and $R_p$ and $v_p$ are the distance and relative velocity of the two galaxies at the closest approach. A merger is assumed to occur if this energy decrement exceeds the kinetic energy at infinity. For identical galaxies, $M_p=M_g$, the critical condition is (the factor of two accounting for the reciprocal perturbations)
\begin{equation}
    2\Delta E = \frac{1}{4}M_gv_{\rm rel}^2\,.
\end{equation}
Energy and angular momentum conservation relate $R_p$ and $v_p$ to $b$ and $v_{\rm rel}$:
\begin{gather}
    0 = \frac{1}{4}M_gv_{\rm rel}^2 - 2\Delta E = \frac{1}{4}M_gv_p^2-\frac{GM_g^2}{R_p}\,,\\
    R_pv_p = bv_{\rm rel}\,,
\end{gather}
which finally implies
\begin{equation}
    \sigma=\pi b^2 = \left(\frac{512\pi^3G^4}{3}\right)^{1/3}M_g^{4/3}R_g^{2/3}v_{\rm rel}^{-8/3}\,. \label{eq:cross_section_galaxy}
\end{equation}
Here, $R_g$ likely positively correlates with $M_g$ \cite{Kravtsov2013}. Numerical simulations demonstrated that the impulse approximation captures the merger behavior even for slow encounters \cite{Dekel1980}. The scaling applies to the BH mass in place of $M_g$ assuming a power-law BH-galaxy mass relation with the exponent close to unity \cite{Kormendy2013}.

Equations~(\ref{eq:cross_section_hardsphere}) and (\ref{eq:cross_section_galaxy}) suggest that the coagulation kernel likely has a positive mass dependence unless $v_{\rm rel}$ strongly increases with mass, which physically underlies our choosing the sum kernel for our model.

\section{Numerical methods for mergers and code test}
\label{app:numerical_mergers}
We implement the numerical merger process in Equation~(\ref{eq:coagulation}) with the forward Euler method, which is first-order accurate in time. Let $M^j$ denote a discrete, log-uniform mass grid indexed by $j$ with a bin width $\Delta\log M$ and $\Phi_{M_\bullet,h}^j$ and $\Phi_{M_\bullet,l}^j$ the heavy- and light-seed BHMF at the center of each mass bin. The discretized equations for the sum kernel $K(M^j,M^k)=B(M^j+M^k)$ are
\begin{gather}
    \Delta \Phi_{M_\bullet,h}^j=D\left[-\Phi_{M_\bullet,h}^j\sum_{k=0}^{k_{\rm max}}(M^j+M^k)(\Phi_{M_\bullet,h}^k+\Phi_{M_\bullet,l}^k) \nonumber\right.\\
    \left.+(M^j)^2\sum_{k=0}^{k_{\rm max}}\left(\frac{\Phi_{M_\bullet,h}^k\Phi_{M_\bullet,h}^n}{M^n} \right.\right. \nonumber \\ \left.\left.+\frac{\Phi_{M_\bullet,h}^k\Phi_{M_\bullet,l}^n}{M^n}+\frac{\Phi_{M_\bullet,l}^k\Phi_{M_\bullet,h}^n}{M^n}\right)\right]\,,\label{eq:discret_phih}
\end{gather}
\begin{gather}
    \Delta \Phi_{M_\bullet,l}^j=D\left[-\Phi_{M_\bullet,l}^j\sum_{k=0}^{k_{\rm max}}(M^j+M^k)(\Phi_{M_\bullet,h}^k+\Phi_{M_\bullet,l}^k) \nonumber\right.\\
    \left.+(M^j)^2\sum_{k=0}^{k_{\rm max}}\frac{\Phi_{M_\bullet,l}^k\Phi_{M_\bullet,l}^n}{M^n}\right]\,,
    \label{eq:discret_phil}
\end{gather}
where $D=B\Delta t\Delta\log M$, $\Delta t$ is the time step, $k_{\rm max}(j)$ satisfies
\begin{equation}
    10^{-\frac{\Delta\log M}{2}}\frac{M^j}{2} \leq M^{k_{\rm max}} < 10^{\frac{\Delta\log M}{2}} \frac{M^j}{2}\,, \nonumber
\end{equation}
and $n(j,k)$ gives the mass index satisfying
\begin{equation}
    10^{-\frac{\Delta\log M}{2}}M^j \leq M^k+M^n < 10^{\frac{\Delta\log M}{2}}M^j\,. \nonumber
\end{equation}
The indices $j,k,n$ correspond to the mass indices of the remnant, the secondary progenitor, and the primary progenitor. In each time step, we first numerically evaluate the heavy-seed BH mass density $\rho_h$ and then obtain $B=1/\tau_m\rho_h$. Then, we update $\Phi_{M_\bullet,h}^j$ and $\Phi_{M_\bullet,l}^j$ using Equations~(\ref{eq:discret_phih}) and (\ref{eq:discret_phil}), i.e., with the forward Euler method. After each time step, we adjust $\Delta t$ (the default value is $\min\{10^{-4}\tau_m, \tau\}$) adaptively such that the numerical heavy-seed BH number density deviates from the theoretical prediction by less than $10^{-5}$. Then, we re-normalize the former to match the latter (cf. \cite{Li2022}, where the mass density is re-normalized instead). We also clear minor negative values in the BHMFs due to numerical errors and reduce $\Delta t$ if any point in the BHMFs drops below $-10^{-9}n_0~\rm dex^{-1}$.

\begin{figure}
    \centering
    \includegraphics[width=0.48\textwidth]{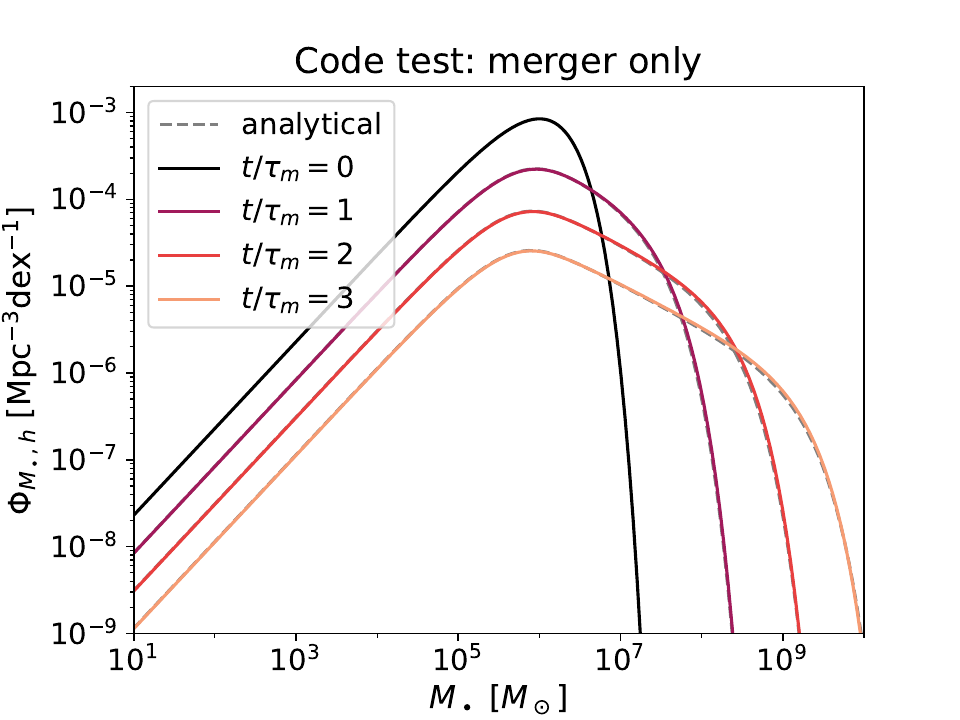}
    \caption{Code test of the sum kernel. The IMF is given by Equation~(\ref{eq:test_IMF}) with $\alpha=0, n_{0h}=10^{-3}~{\rm Mpc^{-3}},$ and $\rho_{0h}=10^3~{M_\odot\,\rm Mpc^{-3}}$. The initial time step is $\Delta t=10^{-3}\tau_m$. The mass function is evolved to the time $t=\tau_m$, $2\tau_m$, and $3\tau_m$ with the forward Euler method. The numerical calculations overall agree with the analytical results.}
    \label{fig:code_test}
\end{figure}

We test our merger implementation using an IMF with a tractable analytical solution. Specifically, we consider the Schechter function \cite{Schechter1976} for the heavy-seed BHs:
\begin{equation}
    \Phi_{M_\bullet,h}(M,t=0)=(\ln10)\Phi^*\left(\frac{M}{M^*}\right)^{\alpha+1} e^{-M/M^*}\,,\label{eq:test_IMF}
\end{equation}
where $\Phi^*$ is a normalization factor, $\alpha$ is a power-law parameter, and $M^*$ is a characteristic mass. We ignore the light-seed BH population here. The initial number and mass densities are given by $n_{0h}=\int_0^\infty\Phi_{M_\bullet,h}(M,t=0)d\log M = \Gamma(\alpha+1)\Phi^*$ and $\rho_{0h}=\int_0^\infty M\Phi_{M_\bullet,h}(M,t=0)d\log M = \Gamma(\alpha+2)\Phi^*M^*$, where $\Gamma(\cdot)$ is the Gamma function. The solution is given by (\cite{Scott1968}, with different notations)
\begin{gather}
    \Phi_{M_\bullet,h}(M,t)=(\ln10)n_{0h}(1-\eta)e^{-[1+\eta/(\alpha+1)]M/M^*}\times \nonumber\\
    \sum_{k=0}^{\infty}\left(\frac{\eta}{\alpha+1}\right)^k\frac{(M/M^*)^{(k+1)(\alpha+1)+k}}{(k+1)!\ \Gamma[(k+1)(\alpha+1)]}\,,
\end{gather}
where $\eta=1-e^{-B\rho_{0h}t}$. Figure~\ref{fig:code_test} shows the test result in the case $\alpha=0$, exhibiting good agreement with the analytical solution. 

\section{Proof of the upper limit for merger rates in the observer's frame}
\label{app:upper_limit}
In this section, we prove the generality of Equations~(\ref{eq:rate_upper_limit}) and (\ref{eq:rate_upper_limit_hl}).

We assume that $z_0$ is the redshift when a significant proportion of the seed BHs have already formed. The value is equal to 20 in this work but may vary in the literature. In general, the observed merger rate earlier than a given redshift $z$ takes the form
\begin{equation}
    N(z) = \int_z^{z_0}\frac{dn}{dz'}F(z')dz'\,,
\end{equation}
where \cite{Haehnelt1994}
\begin{equation}
    F(z') = \frac{1}{1+z'}\frac{dV_C}{dz'}\left|\frac{dz'}{dt}\right|=4\pi c\left[\frac{D_L(z')}{1+z'}\right]^2 \nonumber \,, \label{eq:general_number_reduction}
\end{equation}
and $D_L$ is the luminosity distance. In the equation, $dn/dz'$ denotes the merger-induced BH number reduction over time, subject to the constraint
\begin{equation}
    n_0 > \int_z^{z_0}\frac{dn}{dz'}dz'\,,
\end{equation}
where $n_0$ is the total seed comoving number density. If present, triple/quadruple BH mergers are counted as two/three distinct events.

Cosmology implies that $F(z')$ is positive and monotonically increases with redshift \cite{Haehnelt1994}, so its maximum value in the redshift interval of interest is $F(z_0)$. Therefore,
\begin{equation}
    N(z)<F(z_0)\int_z^{z_0}\frac{dn}{dz'}dz'<F(z_0)n_0\,. \label{eq:general_upper_limit}
\end{equation}
In the limiting case where all BHs merge immediately after birth, such that $dn/dz'\to n_0\delta(z'-z_0)$, Equations~(\ref{eq:rate_upper_limit}) and (\ref{eq:rate_upper_limit_hl}) will be recovered. 

The above proof applies to seeds forming at a given redshift and readily extends to seeding over a redshift range. We only use the number density evolution and cosmology in the proof, so the result is widely applicable. For example, the most abundant heavy-seed scenario in Ref.~\cite{Volonteri2008} predicted $n_0=0.58~\rm Mpc^{-3}$, which, given $z_0=20$, implied $N<2.7\times10^2~\rm yr^{-1}$, consistent with the total merger rates in Ref.~\cite{Klein2016} where the same heavy-seed model was used.

\bibliographystyle{apsrev4-2-my}
\bibliography{bib}
\end{document}